\documentclass[10pt,a4paper]{article}

\date{}

\usepackage{natbib}
\usepackage{amsmath}
\usepackage{amsfonts}
\usepackage{amssymb}
\usepackage{mathtools}
\usepackage{xcolor}
\usepackage{authblk}
\usepackage{verbatim}

\usepackage{soul}
\usepackage{graphicx}
\usepackage[left=2cm,right=2cm,top=2cm,bottom=2cm]{geometry}

\title{Wave-induced drift in third-order deep--water theory}
\author[]{Raphael Stuhlmeier}

\affil[1]{Faculty of Civil \& Environmental Engineering, Technion -- Israel Institute of Technology, Haifa, Israel}

\begin{document}

\newcommand{\K}[1]{\cite[(#1)]{Krasitskii1994}}
\newcommand{\V}[1]{V^{(#1)}}
\newcommand{\A}[1]{A^{(#1)}}
\newcommand{\B}[1]{B^{(#1)}}

\newcommand{\hphi}{\hat{\phi}}
\newcommand{\hpsi}{\hat{\phi}^s}
\newcommand{\hzeta}{\hat{\zeta}}
\newcommand{\bk}{{k}}

\renewcommand{\vec}{\mathbf}
\newcommand\numberthis{\addtocounter{equation}{1}\tag{\theequation}}

\maketitle

\begin{abstract}
The goal of this work is to investigate particle motions beneath unidirectional, deep-water waves up to the third-order in nonlinearity. A particular focus is on the approximation known as Stokes drift, and how it relates to the particle kinematics as computed directly from the particle trajectory mapping. The reduced Hamiltonian formulation of Zakharov and Krasitskii serves as a convenient tool to separate the effects of weak nonlinearity, in particular the appearance of bound harmonics and the mutual corrections to the wave frequencies. By numerical integration of the particle trajectory mappings we are able to compute motions and resulting drift for sea-states with one, two and several harmonics. We find that the classical Stokes drift formulation provides a slight underestimate of the drift at the surface, and a slight overestimate at depth. Incorporating difference harmonic terms into the formulation yields an improved agreement with the drift obtained from nonlinear wave theories, particularly at greater depth. The consequences of this are explored for regular and irregular waves, as well as parametric wave spectra.
\end{abstract}

\section{Introduction}

Waves on the surface of water are a continuing source of fascination, as well as posing an important scientific and engineering challenge, with impacts on such diverse areas as offshore renewables and marine ecology. While we can observe the myriad forms of waves on the water surface, these shapes are only manifestations of what is going on below the surface: a transport of energy and, to a lesser extent, the movement of water particles themselves. Leonardo da Vinci already observed that the wave outruns the water, and early mathematical analyses of wave kinematics seemed to confirm this idea. Work in the early 19th century by Green showed that particle paths in linear deep water waves are approximately circular -- in seeming agreement with the exact, deep-water solution derived by Gerstner -- and the corresponding finite depth particle paths were shown to be approximately elliptical by Airy \citep{Craik2004}.

Nonlinear wave theories began to be explored by Russell, Kelland, and most prominently, Stokes, starting in earnest in the 1840s. It was the work of Stokes that first suggested that particle trajectories in periodic waves in deep water might not be closed, and gave rise to the idea of wave-induced motion now known as Stokes drift. In the intervening years the kinematics of nonlinear waves has received significant attention, and the concept of wave-induced drift has taken a prominent role in our understanding of transport in the ocean. Its consequences are still being explored in the movement of floating microplastics \citep{vanSebille2020} and water-borne bacteria \citep{Ge2012}, and we are coming to develop a fuller picture of particle trajectories in nonlinear wave equations \citep{Carter2020,Curtis2018,Ige2024}.

Nevertheless, much remains unknown from both an experimental and theoretical perspective. Experiments are notoriously difficult, owing to the challenge of comparing a closed flume (with no net mass transport, and with the additional complication of a mechanical wavemaker) with the situation in the (open) ocean. Recent years have seen significant progress, from flume experiments on monochromatic wave trains \citep{Umeyama2012,paprota2018particle} which have been taken to higher and higher steepness \citep{Grue2017} to several recent investigations of particle kinematics for wave groups and complex flows \citep{Bjornestad2021,VandenBremer2019}. 

Meanwhile, theoretical investigations must contend with the fact that the ubiquitous, Eulerian description of fluid motion -- dispensing with particle identity in favour of a simpler treatment of velocity fields -- necessarily obscures certain aspects of kinematics. While these difficulties can be circumvented for steady flows by the use of alternative formulations, including employing stream function and velocity potential as independent variables as originally suggested by Stokes, there is no easy fix for nonsteady, irregular wave fields. While the pre-eminence of Eulerian theory is being challenged by a resurgence of interest in Lagrangian wave mechanics, including higher-order approximate formulations for steady periodic waves \citep{Clamond2007}, standing waves \citep{chen2009third} and, recently, wave packets \citep{abrashkin2018dynamics,pizzo2023role}, the vast majority of studies continue to employ the Eulerian perspective. The non-trivial task of comparing the two approaches \citep{fouques2008comparing} should not be underestimated.

The aim of this work is to explore the theoretical particle trajectories -- and drift -- associated with nonlinear potential flow theories in infinite water depth.  While there are recent mathematical results for the full, nonlinear problem showing that no closed particle paths exist in steady, periodic wave trains \citep{Constantin2006b,Henry2006,Okamoto2012}, these results are not quantitative and are difficult to generalise to nonsteady problems. We will take a `weakly nonlinear' approach, and our starting point will be the Stokes drift formulation which arises in linear (1st order) water wave theory. While a rigorous proof that linear, steady, periodic water waves exhibit a forward drift in the direction of wave propagation was recently found \citep{Constantin2008d,Constantin2008e}, the approximate results due to Stokes are still used in most applications. These were generalised to a spectrum of waves by  \cite{Kenyon1969}, and the resulting formulation has been extensively explored for different spectral shapes \citep{Webb2011}, for its sensitivity to the high-frequency spectral tail \citep{Lenain2020}, and new profiles have been developed for practical implementation \citep{breivik2016stokes}. Many more references can be found in the recent review by \cite{vandenBremer2018}.

As the Stokes drift formulation comes about through a simplification of the linear water-wave theory, the differences between it and the linear theory are of interest. This also raises a natural question: what if we progress from linear theory to a second or third--order theory? We expect the higher-order theories to be more accurate, and to include important effects for steeper waves -- such as the appearance of additional harmonics -- which may have a bearing on comparisons with the Stokes drift. Indeed, these theories may suggest improvements to the Stokes drift derived from linear theory. However, with the exception of recent work focusing on wave groups initiated by van den Bremer and Taylor \citeyearpar{vandenBremer2016} it is difficult to find such comparisons in the literature.

For a successful comparison we need some way to compute the fluid velocity field $\mathbf{u}$, which drives the particle kinematics through the particle trajectory mapping $\mathbf{x}'(t) = \mathbf{u}.$ We need a methodology that can deal with several Fourier modes, to simulate irregular seas, and which is accurate to a desired order of nonlinearity. For more than two Fourier modes the Lagrangian theory is currently limited to second order \citep{Pierson1961}, while important resonant effects in deep water occur only at third order. Consequently, and to increase the accessibility of the results, it seems expedient to work in Eulerian coordinates. Even here several options exist, including the use of traditional perturbation theory to obtain a solution to the water wave problem, which has been extended to third-order by \cite{Madsen2012}. An alternative -- which we employ -- is to use the equivalent Hamiltonian expansion, which is due to \cite{Zakharov1968} and for which the third-order solution has been explored by  \cite{Zhang1999} and \cite{Gao2021}. The Hamiltonian expansion furnishes our velocity field and free surface, and these are the tools we use to examine the particle motions.

The remainder of this paper is structured as follows: in Section \ref{sec:WW problem and Hamiltonian formulation} we introduce the water wave problem and corresponding Hamiltonian. We give more details about the expansion that allows us to extract linear, second, and third-order terms in Section \ref{ssec:Asymptotic analysis and ww}, discuss the neglect of amplitude evolution for our particle kinematics calculations in Section \ref{ssec:Constant magnitude approximation}, and finally show how the free surface and potential are obtained to each order in Section \ref{ssec:Recovery of the third-order solution}. In Section \ref{sec:Monochromatic waves} we consider monochromatic waves, which is the natural setting to introduce Stokes drift, and look at the consequences of higher-order theories on particle trajectories. Section \ref{sec:Bichromatic waves} is devoted to bichromatic waves. These are the simplest type of non-steady wave-field, but because only two harmonics are involved the linear superposition is always periodic. Finally, Section \ref{sec:Multichromatic waves} considers waves involving multiple lowest-order harmonics. A discussion of the work and some conclusions are presented in Section \ref{sec:Conclusions}. Appendix \ref{app:Non-uniqueness of Stokes exp} adds some comments on the non-uniqueness of the monochromatic Stokes wave solution at third-order, while Appendix \ref{app:Coefficients of potential} gives expressions for some coefficients appearing in Section \ref{ssec:Recovery of the third-order solution}. Appendix \ref{app:Amplitude evolution} explores the effects of slow amplitude evolution on particle paths and Stokes drift.

\section{The water wave problem and Hamiltonian formulation}
\label{sec:WW problem and Hamiltonian formulation}

We begin with the inviscid, irrotational and unidirectional water wave problem in Eulerian variables. This can be written in terms of a fluid velocity potential $\phi(x,z,t)$ and free surface $\zeta(x,t)$ as
\begin{subequations}
\begin{align} \label{eq:potential form eq 1}
&\Delta \phi = 0 \text{ on } -h < z < \zeta(x,t), \\ \label{eq:potential form eq 2}
&\zeta_t + \phi_x \zeta_x = \phi_z \text{ on } z = \zeta(x,t), \\ \label{eq:potential form eq 3}
&\phi_t + \frac{1}{2}(\nabla \phi)^2 + g\zeta = 0 \text{ on } z= \zeta(x,t), \\ \label{eq:potential form eq 4}
&\phi_z = 0 \text{ on } z= -h.
\end{align}
\end{subequations}
The advantage of this formulation is that all nonlinearities have been moved to the boundary conditions, which therefore represent the principal challenge. We also note that, provided the water is deeper than about half a typical wavelength, the still-water depth $h$ can be taken to be infinite.\footnote{This assumption is innocuous enough if only one wave with fixed length is considered. However, as we shall see below, the interaction of multiple waves in the sea can give rise to long, bound-harmonics, for which the assumption of infinite depth must be critically reassessed.}  This procedure, using
\begin{equation} \label{eq:potential form eq 4 DW}
\tag{\ref{eq:potential form eq 4}'} \phi_z \rightarrow 0 \text{ as } z \rightarrow \infty
\end{equation}
in place of \eqref{eq:potential form eq 4} simplifies calculations considerably. Despite the restriction to irrotational flow and the neglect of surface tension, compressibility and viscosity, these equations contain a huge range of interesting physics, including many types of wave phenomena. While the techniques we shall employ can be used without alteration for waves with a transverse ($y$) dependence, the additional freedom this engenders means that we defer an exploration of directionally spread waves to future work.

During the late 1960s, efforts to find a Hamiltonian formulation of the water wave problem \eqref{eq:potential form eq 1}--\eqref{eq:potential form eq 4} came to fruition in work of  \cite{Zakharov1968} who found the Hamiltonian
\begin{equation}
\label{eq:Hamiltonian}
H  =  \int \int_{-h}^\zeta  \frac{1}{2} |\nabla \phi|^2 dz \, dx + \int \frac{1}{2}g \zeta^2 dx, 
\end{equation}
which represents the total energy of the fluid. The domain of integration in $x$ can in practice depend on the specifics of the problem, and the lower boundary of the $z$-integral can be chosen as $-\infty$ if \eqref{eq:potential form eq 4 DW} is used in place of \eqref{eq:potential form eq 4}.

The canonical variables for the Hamiltonian formulation are defined at the free surface. They are $\zeta(x,t)$ and $\psi(x,t)=\phi(x,\zeta(x,t),t),$ and the canonical equations  are thus 
\begin{equation}
\frac{\partial \zeta}{\partial t} = \frac{\delta H}{\delta \psi}, \quad \frac{\partial \psi}{\partial t}=-\frac{\delta H}{\delta \zeta}.
\end{equation}
These equations are equivalent to the surface boundary conditions \eqref{eq:potential form eq 2}, \eqref{eq:potential form eq 3}, and details of the calculation can be understood most easily in  \cite{Broer1974}. Additional background can also be found in the recent review by  \cite{Stuhlmeier2024}.

It is worth noting at the outset that any study of kinematics is necessarily somewhat hampered by the Eulerian variables commonly used in wave research. The Eulerian focus on velocity fields as functions of fixed (laboratory) coordinates obscures the motions of fluid particles, which must be laboriously re-derived. Moreover, as we shall see below, an approximate Eulerian theory divorces the motions of the fluid and the free surface, with consequences for our ability to understand the full kinematics. The alternative choice of Lagrangian variables has many advantages \citep{pizzo2023role,blaser2024momentum}, but a general Lagrangian theory of nonlinear, irregular waves has yet to be developed, and the connections to the well-established Eulerian theory remain a topic of active exploration.

\subsection{Asymptotic analysis and the water wave problem}
\label{ssec:Asymptotic analysis and ww}

The water wave problem, whether written as a system of partial differential equations and boundary conditions or using the Hamiltonian reformulation remains mathematically very complex. While the mathematical existence of periodic and solitary wave solutions has been established in ever more general settings (for example, the books by  \cite{Constantin2011k} or  \cite{lannes2013water} offer a relatively recent overview of mathematical progress), much of what is known about qualitative and quantitative properties of  ocean waves rests on an analysis of simplified equations.

These simplified equations are usually obtained through so-called asymptotic analysis or perturbation theory -- a procedure of nondimensionalisation and scaling the equations, followed by the identification of small dimensionless parameters, and an asymptotic expansion of the variables in terms of the small parameters selected. In the case of water waves the most common small parameter is the wave steepness $\epsilon,$ written either as the wave amplitude divided by the wavelength or multiplied by the wavenumber. A very thorough discussion of asymptotic analysis as applied to water waves can be found in the textbook by  \cite{Johnson1997}.

Instead of expanding the water wave equations \eqref{eq:potential form eq 1}--\eqref{eq:potential form eq 4} in this manner (see \cite{Madsen2012} for a detailed, relevant account), it is also possible to expand the Hamiltonian directly. This procedure was carried out by  \cite{Zakharov1968} and, with more detail,  \cite{Krasitskii1994}, and it has the advantage of enforcing structure that might otherwise be lost, such as that resulting equations conserve energy. 
 
The first step in this procedure is to write the canonical equations in terms of the Fourier transformed variables $\hat{\zeta}(k)$ and $\hat{\phi}(k),$ as
\[ \frac{\partial \hat{\zeta}}{\partial t} = \frac{\delta H}{\delta \hat{\psi}^*}, \quad \frac{\partial \hat{\psi}}{\partial t} = - \frac{\delta H}{\delta \hat{\zeta}^*},\]
with $*$ denoting the complex conjugate. Then one defines a new pair of canonical variables $a(k)$ and $ia^*(k)$ related to $\psi$ and $\zeta$ via Fourier transforms as 
\begin{align} \label{eq:Inv-Transf-zeta}
\zeta(x) = \frac{1}{2\pi} \int \left( \frac{q(k)}{2 \omega(k)}\right)^{1/2} \left[ a(k) + a^*(-k)\right] e^{ikx} dk, \\ \label{eq:Inv-Transf-psi}
\psi(x) = \frac{-i}{2\pi} \int \left( \frac{\omega(k)}{2 q(k)}\right)^{1/2} \left[ a(k) - a^*(-k)\right] e^{ikx} dk, 
\end{align}
where 
\begin{equation}
q(k) = |k| \tanh(|k|h),
\end{equation}
$\omega = \sqrt{gq(k)}$ is the frequency (in rad/s), and $g$ is the constant acceleration of gravity (taken to be 9.8 m/s$^2$ in computations). As we will be interested in deep-water waves ($h\rightarrow\infty)$, we note that the dispersion relation simplifies to
\begin{equation} \label{eq: Linear Dispersion Relation}
\omega(\bk)^2 = g | \bk|.
\end{equation}
We will use this in all the computations that follow.

The subsequent steps in the expansion are somewhat lengthy, and are given by  \cite{Krasitskii1994}. At this stage it is important to emphasise only one key point: at lowest order the problem is linear, so the superposition principle holds. This means that the Fourier amplitudes -- which correspond essentially to the magnitudes of the terms $a(k)$ introduced in \eqref{eq:Inv-Transf-zeta}--\eqref{eq:Inv-Transf-psi} -- do not change with time. If gravity is the sole restoring force, the surprising fact is that the Fourier amplitudes do not change with time at the next order, either. Only at third-order in nonlinearity, where resonance between quartets of gravity waves becomes possible, do we obtain an evolution equation. This equation gives the slow evolution of the complex amplitude $b_0=b(k_0,t),$
\begin{equation}  \label{eq:ZE}
i \frac{\partial b_0}{\partial t} = \omega_0 b_0 + \int \tilde{V}^{(2)}_{0123} b_1^* b_2 b_3 \delta_{0+1-2-3} dk_{123},
\end{equation}
and is known as the Zakharov equation, with a kernel term $\tilde{V}^{(2)}$  given in \K{3.5}. We have introduced subscript notation to denote wavenumbers, so that, for example, the delta-function $\delta_{0+1-2-3}$ is a shorthand for $\delta(k_0 + k_1 - k_2 - k_3).$ In principle $k_i$ can be either a scalar or a vector wavenumber, although our computations will deal only with the scalar case for simplicity.

If $b$ is known then the canonical variable $a$ can be recovered via 
\begin{align} \nonumber 
a_0 = b_0 &+ \int \A{1}_{012} b_1 b_2 \delta_{0-1-2} dk_{12} + \int \A{2}_{012} b_1^* b_2 \delta_{0+1-2} dk_{12} + \int \A{3}_{012} b_1^* b_2^* \delta_{0+1+2} dk_{12} \\  \nonumber
&+ \int \B{1}_{0123} b_1 b_2 b_3 \delta_{0-1-2-3} dk_{123} + \int \B{2}_{0123} b_1^* b_2 b_3 \delta_{0+1-2-3} dk_{123} \\  \label{eq:K2.17}
&+ \int \B{3}_{0123} b_1^* b_2^* b_3 \delta_{0+1+2-3} dk_{123} + \int \B{4}_{0123} b_1^* b_2^* b_3^* \delta_{0+1+2+3} dk_{123} + \ldots
\end{align}
where we have written terms up to third order only. This expression is given in \K{2.17}, as are the relevant interaction kernels $A^{(i)}, \, B^{(i)}$. Finally, once $a$ is known it is possible to recover $\zeta$ and $\psi$ up to the desired order by means of \eqref{eq:Inv-Transf-zeta}--\eqref{eq:Inv-Transf-psi}.

\subsection{Constant magnitude approximation}
\label{ssec:Constant magnitude approximation}

The procedure outlined above would seem to suggest that we first need to solve the complicated integro-differential Zakharov equation \eqref{eq:ZE}. In fact, for certain situations of interest we can circumvent this step. The key insight is that the evolution described by the Zakharov equation -- being associated with cubic nonlinearities -- occurs on the slow timescale $\epsilon^{-2} T$, where $\epsilon$ is a typical wave steepness and $T$ is a typical wave period. For processes taking place on faster time-scales, it may be appropriate to consider the complex amplitudes to be constant.

This is indeed the approach we will adopt. There is, however, one important addition: for certain wavenumber combinations the Zakharov equation \eqref{eq:ZE} can be solved explicitly. In particular, for all symmetric quartets which satisfy  $k_0 + k_0 = k_0 + k_0$ or $k_0 + k_1 = k_0 + k_1$ the equation reduces to a system of one or two ordinary differential equations with constant-amplitude solutions, as shown in \cite[Ch.\ 14]{Mei2005}. Such solutions represent nonlinear corrections to the frequency of the waves, and were first found by Stokes for a single wave mode. For arbitrarily many modes, or a continuous spectrum, it is possible to write these corrections in a compact form \citep{Stuhlmeier2019}.

Incorporating these corrections means that the complex amplitudes $b_i(t)$ have constant magnitudes but time-dependent phases
\begin{equation} \label{eq:b_i-t-const-amp-sol} b_i(t) = |b_i(0)| \exp\left( -it \left[ \tilde{V}^{(2)}_{iiii} |b_i(0)|^2 + 2 \sum_{j\neq i} \tilde{V}^{(2)}_{ijij} |b_j(0)|^2 \right] \right). \end{equation}
The terms in the argument of the exponential reduce to $\omega a^2 k^2/2$, the well-known Stokes' correction, when only a single wave-mode is present, or the mutual frequency correction found by \cite{Longuet-Higgins1962d} for two modes. The form \eqref{eq:b_i-t-const-amp-sol} now contains all the mutual dispersion corrections of the cubically nonlinear water wave theory. It has been demonstrated that taking these corrections into account allows one to predict the short-time propagation of laboratory waves with high accuracy without the need to solve for the evolution of the slow-time variables \citep{Stuhlmeier2021,Galvagno2021,Meisner2023}, showing that the constant magnitude approximation is well-justified for such cases. Of course, under specific circumstances -- for example with a sufficiently narrow spectrum, or close to modulation instability -- we may have significant energy transfer \citep{Andrade2023}. However, given our focus on particle paths (with a characteristic time on the order of one period) the amplitude evolution will be considered to be negligible (see also the discussion in \cite{Stuhlmeier2019} and Appendix A therein).

If we sum up the consequences of weak-nonlinearity in the Fourier description\footnote{Only the amplitude-dependence of the dispersion relation is a general property of nonlinear waves. The other effects arise from the viewpoint of Fourier analysis. Solitary waves, for example, do not have a counterpart in linear theory whose ``shape" is modified by the inclusion of higher-order effects.} of surface gravity waves in deep water as 
\begin{enumerate}
\item \, dispersion corrections, \label{item:nl-disp-corr}
\item \, changes to wave shape (bound harmonics), \label{item:nl-wave-shape}
\item \, energy exchange between modes, \label{item:nl-energy-xchange}
\end{enumerate}
we will account for \ref{item:nl-disp-corr}--\ref{item:nl-wave-shape} while neglecting the slow effects \ref{item:nl-energy-xchange}. Appendix \ref{app:Amplitude evolution} provides some further discussion of the neglect of amplitude evolution, along with several examples.

\subsection{Recovery of the third-order solution}
\label{ssec:Recovery of the third-order solution}

The Hamiltonian formulation makes clear how the free surface and potential can be recovered from the canonical variables $a(k)$ and $a^*(k)$ up to a given order. In practice, discrete wavenumbers are considered, which means that the integrals in \eqref{eq:K2.17} can be simplified by the substitution
\begin{equation}
b(k) = \sum_i B_i \delta(k-k_i).
\end{equation}
Here $B_i=B(k_i,t)$ is unknown in principle, and in practice comes from a constant magnitude approximation to \eqref{eq:ZE} as discussed in Section \ref{ssec:Constant magnitude approximation}. 

Because the free surface \eqref{eq:Inv-Transf-zeta} is recovered directly from substitution of \eqref{eq:K2.17}, it is clear that we can distinguish linear, quadratic and cubic terms in $\zeta$ based on the respective terms in \eqref{eq:K2.17}. The same is not true of $\phi$, as we shall see below. 
Taking the inverse Fourier transform yields
\begin{equation} \label{eq:zeta1}
 \zeta_1 = \frac{1}{2 \pi} \sum_i \sqrt{\frac{\omega_i}{2g}} A_i \left( e^{i \xi_i} +  e^{-i \xi_i} \right). 
 \end{equation}
Here we have substituted $B_i = A_i e^{-i\varphi_i(t)}$ with amplitude $A_i \in \mathbb{R}$ and phase $\varphi_i(t),$ and employ the compact notation $\xi_i = k_i x - \Omega_i t + \theta_i.$ Clearly this yields the expected sinusoidal free surface elevation upon defining the amplitude \[ \frac{1}{\pi} \sqrt{\frac{\omega_i}{2g}} A_i = a_i.\] The nonlinear corrected frequency is given by 
\begin{equation}
\label{eq:Omega}
\Omega_i = \omega_i + \tilde{V}^{(2)}_{iiii} |A_i|^2 + 2 \sum_{j\neq i} \tilde{V}^{(2)}_{ijij} |A_j|^2,
\end{equation}
and for unidirectional waves in infinite depth -- which will be our focus -- we can simplify the kernels to 
\[ \tilde{V}^{(2)}_{iiii} = \frac{1}{4 \pi^2} k_i^3 \text{ and } \tilde{V}^{(2)}_{ijij} = \frac{1}{4 \pi^2} k_i k_j \min(k_i,k_j). \]

The quadratic components of the free surface elevation can be obtained in analogous manner, with slightly more in the way of algebra required to resolve all delta-functions. We find 
\begin{align} \nonumber \label{eq:zeta2}
\zeta_2(x,t) = & \frac{1}{2\pi} \sum_{i,j} A_i A_j \left\lbrace \sqrt{\frac{\omega_{i+j}}{2g}} \left( \A{1}_{i+j,i,j} + \A{3}_{-i-j,i,j} \right) \left[ e^{i(\xi_i + \xi_j)}  + e^{-i(\xi_i + \xi_j)}  \right]  \right. \\ 
& \left. + \sqrt{\frac{\omega_{i-j}}{2g}} \A{2}_{j-i,i,j} \left[  e^{i(\xi_j-\xi_i)} + e^{i(\xi_i-\xi_j)} \right] \right\rbrace. 
\end{align}
The shorthand notation $\omega_{i\pm j}$ is used to denote $\omega(k_i \pm k_j) = \sqrt{g |k_i \pm k_j|}.$

We can write the cubic components of the free surface elevation as 
\begin{align} \nonumber \label{eq:zeta3}
\zeta_3(x,t) = & \frac{1}{2 \pi} \sum_{i,j,k} A_i A_j A_k \left\lbrace \sqrt{\frac{\omega_{i+j+k}}{2g}} \left( \B{1}_{i+j+k,i,j,k} + \B{4}_{-i-j-k,i,j,k} \right) \left[ e^{i(\xi_i + \xi_j + \xi_k)}  \right. \right. \\  \nonumber
& \left. + e^{-i(\xi_i + \xi_j + \xi_k)}  \right] + \sqrt{\frac{\omega_{i-j-k}}{2g}} \B{2}_{j+k-i,i,j,k} \left[ e^{i(\xi_j+\xi_k-\xi_i)}  \right.   \left. \left. + e^{i(\xi_i -\xi_j - \xi_k)} \right] \right. \\
& \left. + \sqrt{\frac{\omega_{i+j-k}}{2g}} \B{3}_{k-i-j,i,j,k} \left[ e^{i(\xi_k-\xi_i-\xi_j)}  + e^{i(\xi_i+\xi_j-\xi_k)}  \right] \right\rbrace.
\end{align}
The velocity potential at the free surface, $\psi$, is obtained from \eqref{eq:Inv-Transf-psi} via an identical procedure.

\begin{figure}
\centering
\includegraphics[width=0.8\linewidth]{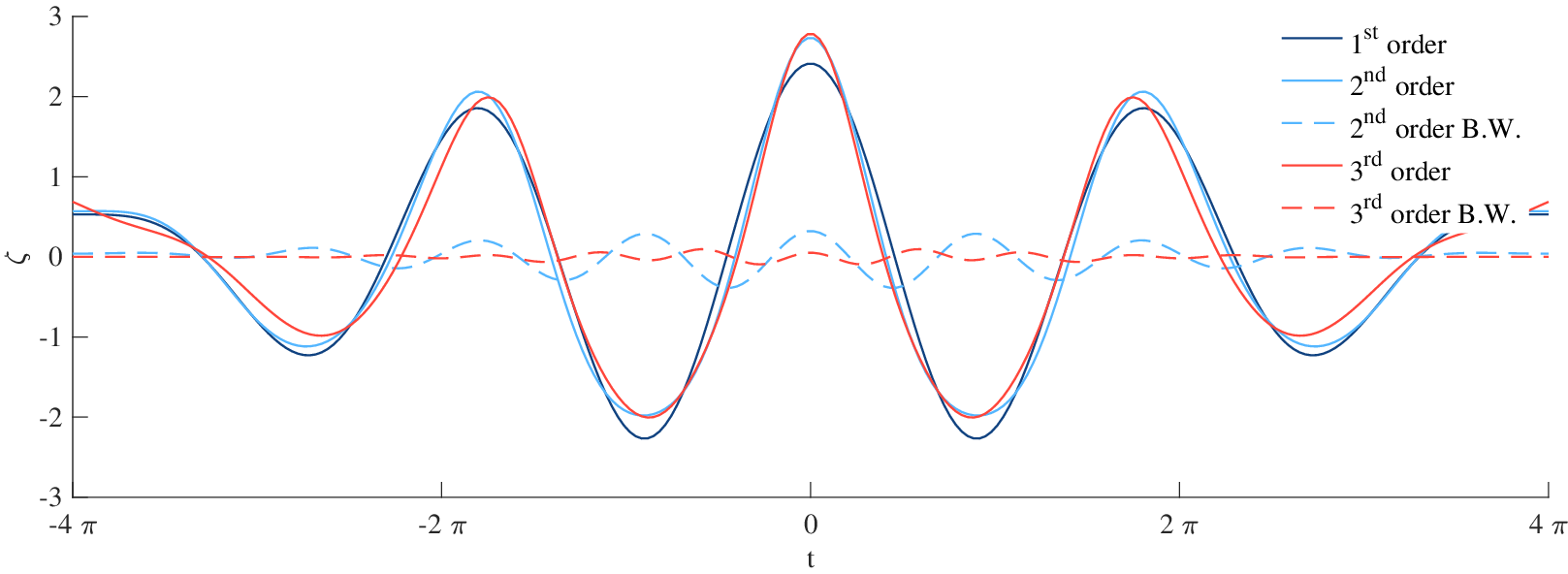} 
\caption{Free surface of a bichromatic wave train with linear frequencies $\omega_1=1$ rad/s and $\omega_2=1.25$ rad/s and wave slopes $\epsilon_1=\epsilon_2=0.15,$ showing 1st order (linear), 2nd order, and 3rd order solutions and their bound wave (B.W.) constituents. Note the phase shift appearing at 3rd order due to the frequency correction \eqref{eq:Omega-nonlin-bichrom-short-wave-1}--\eqref{eq:Omega-nonlin-bichrom-long-wave-2}.}
\label{fig:Bichromatic-free-surface-constituents}
\end{figure}

The contributions of the second-order and third-order free surface elevation change only the wave shape. The resulting harmonics of the form $\exp(i(\xi_i \pm \xi_j))$ behave as waves with wavenumber $k_i \pm k_j$ and frequency $\Omega_i \pm \Omega_j,$ but because these sums and differences do not satisfy the dispersion relation they are referred to as ``bound waves", in contrast to the free waves of \eqref{eq:zeta1}. The impact of these bound waves on sharpening the crests and flattening the troughs is shown in Figure \ref{fig:Bichromatic-free-surface-constituents} for a bichromatic sea, together with the effect of the third-order frequency correction \eqref{eq:Omega} on the wave phases. We also note the important effect of the difference harmonic terms $k_i-k_j$ in creating a ``set-down" below the crest, see \cite{VandenBremer2019}.

\subsubsection{Recovery of the bulk potential}

The reformulation of the problem in surface variables is very convenient from many perspectives, but is something of a disadvantage when we wish to recover information about the bulk fluid below the free surface. To obtain this we start with the general solution of the Laplace equation which decays for infinite depth 
\begin{equation}
\phi(x,z) = \frac{1}{2\pi} \int \phi(k) e^{kz} e^{ikx} dk, \quad \phi(k) = \phi^*(-k),
\end{equation}
the exact analogue of \cite[(4.1)]{Krasitskii1994}.

We relate the bulk potential $\phi$ and its surface trace $\psi$ to one another in terms of an expansion.  Taking the problem in infinite depth, we need to solve 
\begin{align}
&\Delta \phi = 0, \\
&\phi = \psi \text{ on } z = \zeta, \label{eq:bulk-potential-bc}\\
&\phi_z \rightarrow 0 \text{ as } z \rightarrow -\infty,
\end{align} 
which we do by assuming the free surface elevation is small ($\zeta \rightarrow \epsilon \zeta$), and then inserting this expansion into the boundary condition \eqref{eq:bulk-potential-bc}, so that 
\[ \phi(\zeta) = \phi(0) + \epsilon \zeta \phi_z(0) + \frac{\epsilon^2 \zeta^2}{2} \phi_zz(0) + \ldots. \]
This gives rise to a hierarchy of problems in physical space, whose successive solution yields
\begin{equation} \label{eq:phi-FT-expansion}
\hphi(k,z,t) = e^{|k|z} \left[ \hpsi - \frac{1}{2 \pi} \int \hzeta_1 \hpsi_2 |k_2| \delta_{0-1-2} dk_{12} - \int D_{0123} \hpsi_1 \hzeta_2 \hzeta_3 \delta_{0-1-2-3} dk_{123} + \ldots \right]
\end{equation}
with
\begin{equation} \label{eq:DNO_inversion_d} D_{0123} = -\frac{1}{4} |k_1| \left( 2 |k_1| - |k_1 + k_2| - |k_1 + k_3| - |k_0-k_3| - |k_0-k_2| \right). 
\end{equation}
With $\hat{\psi}$ and $\hat{\zeta}$ known from the formulation in \eqref{eq:Inv-Transf-zeta}--\eqref{eq:Inv-Transf-psi} we can now recover the bulk potential at each order desired. This procedure is also detailed in \cite[below (4.7)]{Janssen2004}, \cite[p.15]{Krasitskii1994} and \cite[(1.8)]{Zakharov1968}, although care should be taken in the symmetrisation of \eqref{eq:DNO_inversion_d}.

The linear contribution consists simply of the first term in \eqref{eq:phi-FT-expansion}
\begin{equation}
\phi_1(x,z,t)  =  \frac{1}{\pi} \sum_i A_i e^{|k_i|z} \sqrt{\frac{g}{2\omega_i}} \sin(\xi_i).
\end{equation}
The quadratic contribution is 
\begin{align*} \nonumber
\phi_2(x,z,t) & =  \sum_{i,j} A_i A_j  \left[ \mathcal{C}^{(2)}_{i+j} \sin(\xi_i + \xi_j) e^{|k_i+k_j|z} + \mathcal{C}^{(2)}_{i-j} \sin(\xi_i - \xi_j) e^{|k_i-k_j|z} \right] 
\end{align*}
where we need to employ the linear and quadratic parts of $\hpsi$ and $\hzeta$ (and products thereof) in the formulation. The coefficients $\mathcal{C}^{(2)}$ are given in Appendix \ref{app:Coefficients of potential}.
Up to second order, for unidirectional waves in deep water the potential reduces to the simple expression
\begin{equation} \label{eq:phi-to-second-order}
\phi = \sum_j \frac{a_j g}{\omega_j} e^{|k_j|z} \sin ( k_j x - \omega_j t) - \sum_{i>j} \omega_{i} a_i a_j \sin(\xi_{i} - \xi_{j} ) e^{|k_i - k_j| z},
\end{equation}
which follows also from \cite{Dalzell1999}.

The cubic contribution to the bulk potential is
\begin{align*}
 \phi_3(x,z,t) &= \sum_{ijk} A_i A_j A_k \left\lbrace \mathcal{C}^{(3)}_{i+j+k} \sin(\xi_i+\xi_j+\xi_k) e^{|k_i+k_j+k_k|z} \right. \\
 &+ \mathcal{C}^{(3)}_{i-j-k} \sin(\xi_i-\xi_j-\xi_k) e^{|k_i-k_j-k_k|z}  \left.  + \mathcal{C}^{(3)}_{i+j-k} \sin(\xi_i + \xi_j - \xi_k) e^{|k_i+k_j-k_k|z} \right. \\
 &+ \left. \mathcal{C}^{(3)}_{i-j+k} \sin(\xi_i-\xi_j+\xi_k) e^{|k_i-k_j+k_k|z} \right\rbrace. \numberthis \label{eq:phi-to-third-order}
\end{align*}
The coefficients $\mathcal{C}^{(3)}$ are given in Appendix \ref{app:Coefficients of potential}.
Unfortunately efforts to find a compact simplification the third order have failed except in the case of a single monochromatic wave, where the analytical expression 
\[ \phi_3  = -\frac{a^3 \omega k}{4} e^{kz} \sin \xi \]
is recovered (see \cite{Gao2021} and Appendix \ref{app:Non-uniqueness of Stokes exp}).

It should be emphasised that the bulk potential thus obtained is equivalent (except for the non-uniqueness at a given order mentioned in Appendix \ref{app:Non-uniqueness of Stokes exp}) to the potential found through direct perturbation expansion of the governing equations \eqref{eq:potential form eq 1}--\eqref{eq:potential form eq 4 DW}. In particular, as we shall see below, this means that the fluid domain at each order is the half-space $\{ (x,z) \mid x\in \mathbb{R}, z\leq 0 \},$ owing to the transfer of the boundary conditions.

\section{Monochromatic waves}
\label{sec:Monochromatic waves}

The mathematically simplest type of wave motion, and the one about which we know the most, is that of steady, periodic waves known as Stokes waves. For such waves, symmetric about a crest and propagating without change of form, it can be proven that no closed particle trajectories exist \citep{Constantin2006b}, and explicit calculations of the surface drift have been given by \cite{Longuet-Higgins1987}, up to and including the wave of greatest height. This wealth of prior results means that we do not strictly need the Hamiltonian expansion developed in Section \ref{sec:WW problem and Hamiltonian formulation}; however we shall attempt to situate the 3rd order theory and its predictions among other results for monochromatic waves.

Because the wave-form is steady and the fluid motion is periodic and confined to the $(x,z)$-plane, a variety of powerful mathematical techniques exist for tackling the problem, including the use of the velocity potential and stream function as independent variables \citep{Stokes2009}. Moreover, the convergence of perturbative solutions for such waves has been established since the work of \cite{Levi-Civita1925}, and in practice such solutions have been computed to extremely high order \citep{schwartz1974computer}. 

Thus, if we fix the wave length (or wavenumber) and the wave height $H,$ the successive inclusion of higher order terms brings us ever closer to the exact solution. This appears to be true also when we employ the equivalent Lagrangian formulation of fluid mechanics \citep{Clamond2007}, albeit without the accompanying analytical theory proving convergence. Some care must be taken in the selection of expansion parameter (in which the ``order" is measured), as discussed by \cite{cokelet1977steep} and \cite{Fenton1985}, about which more will be said below.

Although these steady, periodic waves are theoretically exceptional and experimentally difficult to realise, the wealth of accumulated understanding makes them an ideal starting point, and allows for comparisons that will stand us in good stead for later, unsteady problems. We start by reproducing the textbook monochromatic solution \citep{Dean1991} to the linearised governing equations \eqref{eq:potential form eq 1}--\eqref{eq:potential form eq 4 DW} in the Eulerian description:
\begin{align} \label{eq:lin-solution}
& \zeta_1 = a \cos(\xi), \quad \phi_1 = \frac{a\omega}{k}  e^{kz} \sin(\xi),
\end{align}
where $\xi=kx-\omega t$ and 
\begin{equation}
\omega^2 = g|k|
\end{equation}
is the linear dispersion relation. Here $z\leq 0$ and $x \in \mathbb{R}.$ If the gradient of $\phi_1$ from \eqref{eq:lin-solution} is evaluated to yield the velocity field, it is evident that the average over a wave period $T$ -- called the \textit{Eulerian mean velocity}, and whose horizontal component is denoted by ${u_E}$ -- vanishes at every depth $z\leq 0$. 

Just because the average velocity measured by a fixed sensor vanishes does not mean that a particle released at the sensor location returns there after one wave period. The averaged horizontal velocity of such a particle will be called the \textit{Lagrangian mean velocity} $u_L$, and the difference between $u_E$ and $u_L$ is then known as the \textit{Stokes drift} $u_S$, such that $u_L-u_E=u_S.$ 

It is worth mentioning that the Eulerian and Lagrangian averages are not, generally, taken over the same time interval. Following \cite{Longuet-Higgins1987}, we define the Lagrangian period $T_L$ as 
\begin{equation}
T_L = \frac{T_E}{1-{u_S}/{c_p}},
\end{equation}
where $T_E=2\pi/\omega$ is the Eulerian period, $c_p=\omega/k$ is the phase speed of the wave, and $u_s$ is the Stokes drift at a given depth. Clearly the two periods coincide only when the Stokes drift vanishes. If we define, following \cite{Grue2020}, a complete particle loop as the location and time when a particle path, corrected for mean drift distance, begins to repeat itself, then the period of this loop is the Lagrangian period.

To obtain the trajectory of a particle from our Eulerian description, we resort to the system of ordinary differential equations 
\begin{align} \label{eq:particle-trajectory-mapping}
\vec{x}'(t) = \nabla \phi(x,z,t)
\end{align}
known as the \textit{particle trajectory mapping}. It is worth noting that this system has only a tenuous connection to the linear (or, later, weakly nonlinear) theory, but instead attempts to restore the link between Lagrangian and Eulerian descriptions of the fluid motion. In addition, it is an unwelcome surprise that inserting the linear potential \eqref{eq:lin-solution} into \eqref{eq:particle-trajectory-mapping} yields a nonlinear system of differential equations
\begin{align} \label{eq:lin-part-path-ODE-x}
x'(t) = a \omega e^{kz} \cos(\xi),\\ \label{eq:lin-part-path-ODE-z}
z'(t) = a \omega e^{kz} \sin(\xi).
\end{align}
While this system cannot be solved analytically it has been shown rigorously that its particle trajectories are not closed \citep{Constantin2008d}, implying the existence of a Lagrangian drift. The result is qualitative, and an analytical treatment that yields the drift quantitatively even in this simple case appears out of reach.

In lieu of this, two alternatives remain: approximation or numerical solution. The former approach is found in most textbooks on water waves, and begins with a Taylor expansion of the fluid velocity field about an initial position $\vec{x}_0=(x_0,z_0)$. Keeping only the lowest order terms in the Taylor expansion yields, upon integration, the circular particle trajectories first found by Green in 1839 \citep{Craik2004}, for which it is clear that $u_E=u_L=0$ and so that $T_E = T_L = 2\pi/\omega.$ Inserting that solution into the second-order Taylor expansion yields another explicit system of ODEs
\begin{align} \label{eq:x'-2nd-order-Taylor}
x'(t) &= a \omega e^{kz_0} \cos(\xi_0) - a^2 k \omega e^{2kz_0} \cos(\omega t) + a^2 k \omega e^{2kz_0},\\ 
z'(t) &= a \omega e^{kz_0} \sin(\xi_0) + a^2 k \omega e^{2kz_0} \sin(\omega t), \label{eq:z'-2nd-order-Taylor}
\end{align}
where $\xi_0 = kx_0 - \omega t$. Strictly speaking the time-integration of this system yields a Lagrangian displacement, whose average is a Lagrangian mean velocity $u_L$. It is immediate that this mean velocity is determined by the secular term obtained from integrating \eqref{eq:x'-2nd-order-Taylor}, and since $u_E \equiv 0$ it is both conventional and appropriate to call this term
\begin{equation} \label{eq:lin-monoch-Stokes-drift}
u_S = a^2 k^2  c_p  e^{2kz_0},
\end{equation}
the Stokes drift. This Stokes drift is nominally a second order -- and therefore nonlinear -- quantity in the small wave steepness $ak$, but is derived from linear theory \eqref{eq:lin-solution}, a fact which is also recalled in many textbooks, such as  \cite{Dean1991}. In fact, it is an approximation of the linear particle trajectories, obtained under the assumption that the displacement $\Delta \vec{x}$ from the original position $\vec{x}_0$ is small.

\begin{figure}[h]
\centering
\includegraphics[width=\linewidth]{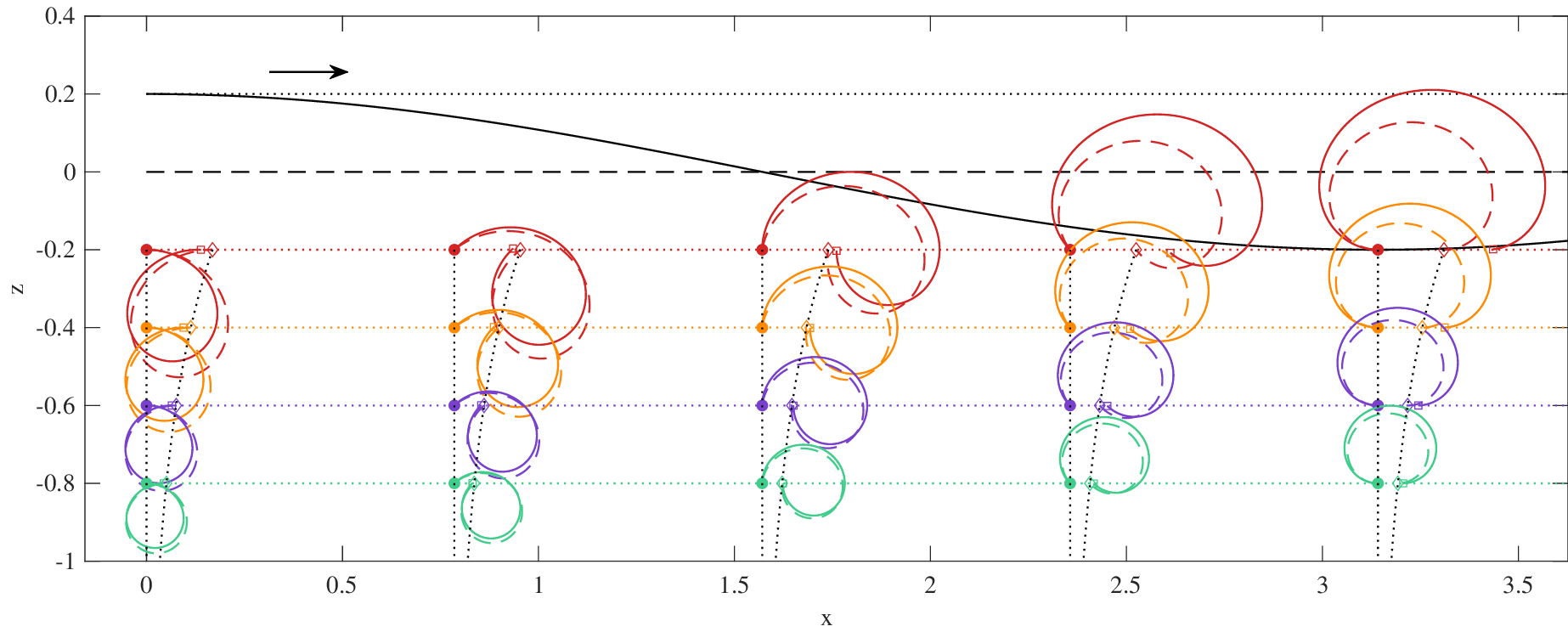} 
\caption{A linear, monochromatic wave profile with $k=1$ 1/m and $H=0.4$ m propagating in the positive $x$-direction. Particle paths are denoted by coloured curves, with solid curves denoting the explicit integration of the particle trajectory ODEs \eqref{eq:lin-part-path-ODE-x}--\eqref{eq:lin-part-path-ODE-z} and dashed curves the circular trajectories with Stokes drift \eqref{eq:lin-monoch-Stokes-drift} obtained from the approximate linear theory. The Stokes drift is shown in thin, dotted curves connecting the initial position (filled circle) and the final position (diamond) obtained from \eqref{eq:x'-2nd-order-Taylor}--\eqref{eq:z'-2nd-order-Taylor}.}
\label{fig:Linear-Monoch-Stokes-Drift}
\end{figure}

Figure \ref{fig:Linear-Monoch-Stokes-Drift} compares the solution of the linear particle trajectory mapping \eqref{eq:lin-part-path-ODE-x}--\eqref{eq:lin-part-path-ODE-z} with that of its approximation \eqref{eq:x'-2nd-order-Taylor}--\eqref{eq:z'-2nd-order-Taylor}, where a rather large value of $ak=0.2$ is used for illustration.  In the approximation \eqref{eq:x'-2nd-order-Taylor}--\eqref{eq:z'-2nd-order-Taylor}, shown in dashed curves in Figure \ref{fig:Linear-Monoch-Stokes-Drift}, after one Eulerian period each particle has moved a uniform amount $u_S T_E$ to the right (diamond markers). By contrast, directly integrating \eqref{eq:lin-part-path-ODE-x}--\eqref{eq:lin-part-path-ODE-z} yields a forward drift that depends on the initial position (square markers), and which is sometimes smaller and sometimes larger than the Stokes drift \eqref{eq:lin-monoch-Stokes-drift}. This dependence of the drift on the phase is also observed experimentally \citep{Grue2020}.

Figure \ref{fig:Linear-Monoch-Stokes-Drift} also emphasises a crucial blind-spot of the linear theory: the fluid velocity field is defined only below $z=0.$ Thus even a particle starting initially at trough level -- such as those on the right side of Figure \ref{fig:Linear-Monoch-Stokes-Drift} -- will enter a region of the $(x,z)$-plane where the velocities are undefined (indeed, particle paths so-calculated may exceed the crest height, as can be seen in the red trajectory on the far right; we should not make too much of this fact, since the free surface in linear (Eulerian) theory is not a streamline of the flow, nor is it composed of fluid particles). An evaluation of the velocity field above $z=0$ amounts to an extrapolation of the linear theory, a procedure which is known to overestimate velocities near the surface \citep{johannessen2010calculations}, and so caution is required when evaluating Lagrangian flow properties from (approximate) Eulerian solutions between crest and trough levels.

Our approach will therefore be as follows: integration of the particle trajectory mapping \eqref{eq:particle-trajectory-mapping} yields the Lagrangian drift and thereby the Lagrangian mean velocity. This value is phase-dependent, which means we average the Lagrangian mean velocities for initial positions covering one wavelength. The integration can be carried out provided we remain in the region $z \leq 0$ where the Eulerian velocity field $\nabla \phi$ is defined, and the value for $u_L$ so obtained can be compared with the leading-order approximation to $u_S$ given in \eqref{eq:lin-monoch-Stokes-drift}. We will not apply this procedure at the surface, due to the aforementioned issue with the domain of definition of the velocity field.

\subsection{Lagrangian drift with depth}

The depth-dependent Lagrangian drift can be obtained from integrating the particle trajectory mapping, provided particles do not cross the still-water level $z=0$ -- otherwise, as mentioned above, extrapolating the terms $\exp(kz)$ can lead to spuriously high velocities or even divergence of the trajectories. Because the forward drift of a particle is phase-dependent, this means averaging over the phases of particles lying at an initial level $z_0.$ In addition to the linear solution \eqref{eq:lin-solution}, we will employ Eulerian solutions
\begin{align}
\phi = \frac{a \omega}{k}e^{kz} \sin \xi + \frac{a^4 k^2 \omega }{2} e^{2kz} \sin 2 \xi + \frac{a^5 k^3 \omega}{12}e^{3 kz} \sin 3 \xi + \ldots \label{eq:mono-pot-5th-order}
\end{align}
up to fifth order (2nd and 3rd order may be obtained from Section \ref{ssec:Recovery of the third-order solution}, see also Appendix \ref{app:Non-uniqueness of Stokes exp}, 4th and 5th order are given for Stokes waves by \cite{zhao2022stokes}), which can be readily inserted into the right-hand side of the particle trajectory mapping \eqref{eq:particle-trajectory-mapping}. From \eqref{eq:mono-pot-5th-order} it can be noted that $u_E$ is zero at all orders, so that the Lagrangian drift $u_L$ is equal to the Stokes drift $u_S,$ whose leading order constituent is given in \eqref{eq:lin-monoch-Stokes-drift}.

\begin{figure}[h]
\centering
\includegraphics[width=\linewidth]{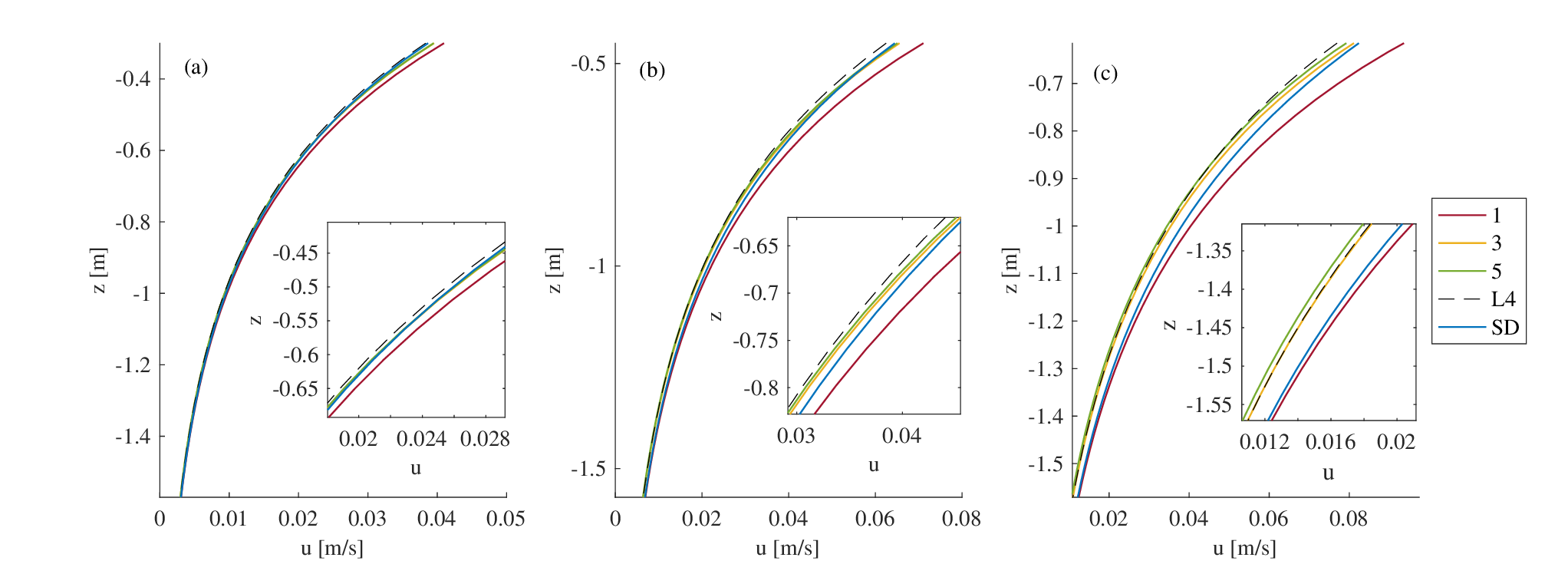} 
\caption{Comparison of horizontal drift velocities with depth beneath monochromatic waves with $k=1$ 1/m and (a) $H=0.3$, (b) $H=0.45$ and (c) $H=0.6$. Particle trajectories are obtained at 1st, 3rd and 5th order from integration of the particle trajectory ODEs and yield Lagrangian drift velocities (labels 1, 3, 5). These are compared with the Stokes drift approximation \eqref{eq:lin-monoch-Stokes-drift} (SD) and the fourth order Lagrangian solution \cite{blaser2025increased} (L4).}
\label{fig:Depth-Stokes-drift-monochromatic}
\end{figure}

In Figure \ref{fig:Depth-Stokes-drift-monochromatic}, 50 equally-spaced initial conditions covering the wave phase are used at each depth to compute the Lagrangian drift. We compare results using 1st, 3rd and 5th order velocity fields from \eqref{eq:mono-pot-5th-order}, as well as a 4th order Stokes drift from the Lagrangian approach of \cite{blaser2025increased} (see also \cite{Clamond2007}), and which we shall return to for unsteady cases below. The vertical axes show depth below $z=0$, while the horizontal axes show drift velocity $u$;  for labels 1, 2, 3 and L4 this is formally $u_L$ (although $u_E=0$ means $u_L=u_S$) while label SD shows the leading order approximation to $u_S$ given by \eqref{eq:lin-monoch-Stokes-drift}.

It is noteworthy that the integration of the 1st order velocity field (red curves, label 1) gives a significant overestimate of the Lagrangian drift with depth, while the approximation \eqref{eq:lin-monoch-Stokes-drift} based on that same velocity field (blue curves, label SD) gives much better agreement with both the 4th order Lagrangian result and higher-order Eulerian velocity fields. Only for high steepness $kH/2=0.225$ and $0.3$ shown in panels (b)--(c) is the approximate Stokes drift noticeably different from the higher-order solutions, although its asymptotics at large depth hew close to the 1st order theory, while the higher-order solutions have a somewhat different behaviour at large depth (see inset in panel (c)). Nevertheless, these results bear out the fact that -- for steady, monochromatic waves -- the Stokes drift is very well approximated by the leading order (quadratic) contribution \eqref{eq:lin-monoch-Stokes-drift}.

\subsection{Lagrangian drift at the surface}
\label{ssec:Mono-Lag-drift-at-surface}

It appears that the most natural way to obtain surface drift values from an approximate Eulerian theory is by solving for the horizontal particle displacement from
\begin{equation} \label{eq:particle-trajectory-map-surface}
x'(t)=u(x,\zeta,t) 
\end{equation} (see \cite{Grue2020}, who employ this to 2nd order to compute the Lagrangian period) where the right-hand side can be approximated as 
\[ u(x,\zeta,t) = u(x,0,t) + \zeta u_z(x,0,t) + \frac{\zeta^2}{2} u_{zz}(x,0,t) + \ldots,\]
making use of the expansion of the free surface 
\begin{align} \nonumber
\zeta =& a \cos \xi \left[ 1 + \frac{1}{8} a^2 k^2 + \frac{121}{192} a^4 k^4 \right] + a \cos 2 \xi \left[ \frac{1}{2}ak + \frac{5}{6}a^3 k^3 \right] \\
& + a \cos 3 \xi \left[ \frac{3}{8}a^2 k^2 + \frac{171}{128}a^4 k^4 \right] + a \cos 4 \xi \left[\frac{1}{3} a^3 k^3 \right]+a \cos 5 \xi \left[ \frac{125}{384}a^4 k^4 \right] + \ldots \label{eq:mono-fs-5th-order}
\end{align}
together with the potential given in \eqref{eq:mono-pot-5th-order}.

\begin{figure}[h]
\centering
\includegraphics[width=\linewidth]{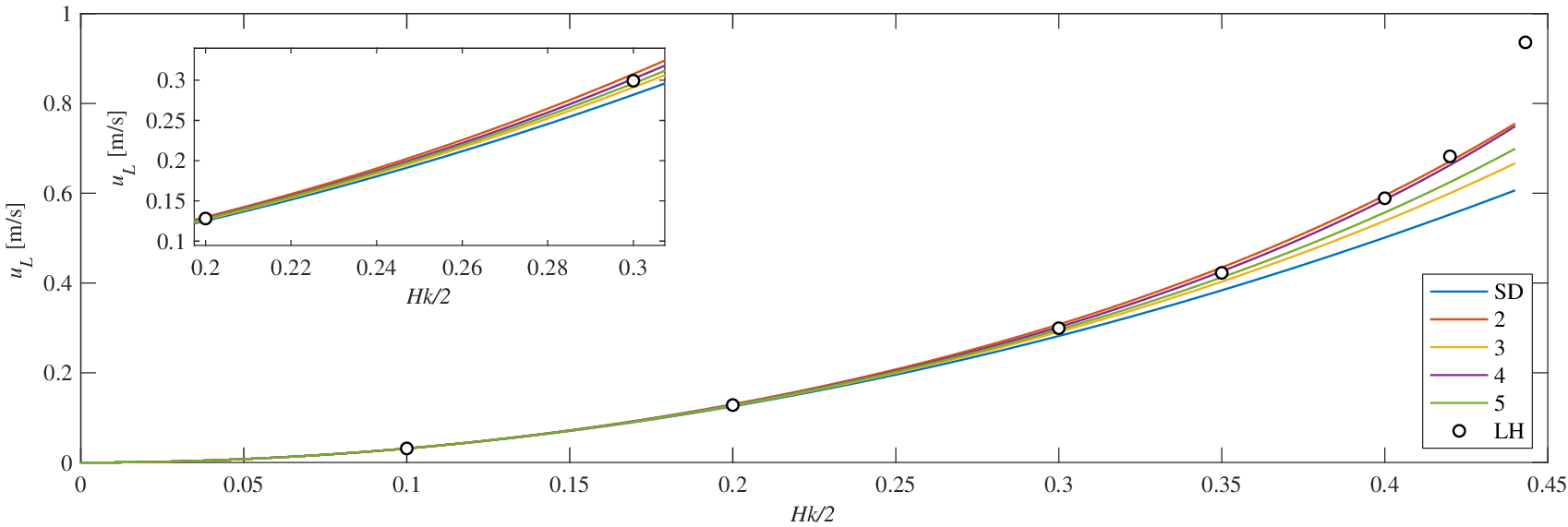} 
\caption{Plots of Lagrangian surface drift velocity $u_L$ for monochromatic waves of varying steepness $Hk/2$, using the surface velocity from 2nd to 5th order (2--5), compared with the exact solution of Longuet-Higgins (LH) and the approximate Stokes drift $u_S$ \eqref{eq:lin-monoch-Stokes-drift} (SD).}
\label{fig:Surface-Stokes-drift-monochromatic}
\end{figure}

Results are shown in Figure \ref{fig:Surface-Stokes-drift-monochromatic}, which compares the formula \eqref{eq:lin-monoch-Stokes-drift} evaluated at $z_0=0$ (label SD) with solutions of \eqref{eq:particle-trajectory-map-surface} up to 5th order, as well as the exact surface drift obtained in \cite[Fig.\ 2]{Longuet-Higgins1987} (label LH). Taking $k=1$ 1/m, the horizontal axis denotes half the actual crest-to-trough height -- not the leading-order amplitude $a$ which is used in the perturbation expansion -- which must be adjusted every odd order to obtain a desired value of $H$ (see discussion in Appendix \ref{app:Non-uniqueness of Stokes exp}). Without such an adjustment the wave height grows with the order of the expansion, as does the surface drift. 

Figure \ref{fig:Surface-Stokes-drift-monochromatic} shows that for waves of low steepness (below $Hk/2=0.2$) the differences in the formulations are essentially negligible. The Stokes drift formula \eqref{eq:lin-monoch-Stokes-drift} gives a drift that is slightly too small (as shown by \cite{Longuet-Higgins1987} and complementary work using the Lagrangian formulation, such as \cite{Clamond2007}), but higher-order theories provide good agreement up to very high steepness (though the trend clearly does not continue to Longuet-Higgins' steepest wave with $kH/2=0.44316$). We note that dispersion corrections appearing at the 3rd and 5th order have the effect of decreasing surface drift slightly, which we comment upon below. The seemingly excellent agreement between the 2nd order approximation and the exact solution at slopes above $Hk/2=0.35$ should be considered accidental.

\subsection{Remarks on the higher-order contributions}

In the formulation of the velocity potential used here (corresponding to \eqref{eq:App-Sol-1a} in Appendix \ref{app:Non-uniqueness of Stokes exp}) the only difference between the 1st and 3rd order is only the inclusion of nonlinear dispersion, so that 3rd order waves travel with a characteristic coordinate $\xi = kx - \Omega t$ for 
\begin{equation} \label{eq:Monochr-Stokes-correction}
 \Omega = \omega(k) \left(1 + \frac{a^2 k^2}{2} \right).
 \footnote{Note $a$ here is related to the wave height $H_1$ as described in \eqref{eq:App-a-H-rel-1-2}.}
\end{equation}
No additional harmonics appear until 4th order, so the difference between curves 1 and 3 in Figure \ref{fig:Depth-Stokes-drift-monochromatic} is entirely a consequence of this dispersion correction. Indeed, the procedure used to derive \eqref{eq:lin-monoch-Stokes-drift} can be applied without alteration to the 3rd order potential $\phi$ i.e.\ by calculating $\overline{\Delta \vec{x}^\intercal \nabla \phi}$ , yielding
\begin{equation} \label{eq:NL-Stokes-drift-monochrom}
{u_S} = \frac{a^2 \omega^2 k e^{2kz_0}}{\Omega} = \frac{2 a^2 \omega k e^{2kz_0}}{2+a^2 k^2}.
\end{equation}
Because $\Omega > \omega$ it is clear that the drift velocity \eqref{eq:NL-Stokes-drift-monochrom} is generally smaller than \eqref{eq:lin-monoch-Stokes-drift}, as observed also from integrating the particle trajectory mapping. 

We can summarise the results for monochromatic waves as follows: the approximate Stokes drift formula \eqref{eq:lin-monoch-Stokes-drift} gives results that are somewhat too small at the surface, and somewhat too large at depth. The differences for waves of moderate steepness, however, are not sizeable, as borne out also by experimental work \citep{paprota2018particle}.  In our description, the only effects of nonlinearity are in adjusting the frequency (which has consequences for the entire flow) and in the appearance of higher-harmonics (whose effect is confined close to the surface due to the $\exp(nkz)$-terms, see \eqref{eq:mono-pot-5th-order}). We shall now see how these effects appear in unsteady (bichromatic) wave trains.

\section{Bichromatic waves}
\label{sec:Bichromatic waves}

The simplest ``irregular" waves, corresponding to unsteady flows, are those that consist of two harmonics $k_1$ and $k_2$ at first order, and are therefore sometimes termed ``bichromatic" waves. The second order solution then contains sum and difference terms of any two of these, while the third order contains sum and difference terms of any three, with an attendant increase in complexity. Up to second order the solution is contained in the work of  \cite{Dalzell1999}, while the third order solution for bichromatic waves appears first in the work of  \cite{Madsen2006}. We make use of the equivalent formulation of Section \ref{ssec:Recovery of the third-order solution}.

Up to second order the velocity potential can be written as the remarkably simple expression
\begin{align}
\label{eq:Second_order_bichromatic}
\phi(x,z,t) = \frac{a_1 g}{\omega_1} e^{k_1 z} \sin(\xi_1) + \frac{a_2 g}{\omega_2} e^{k_2 z} \sin(\xi_2) - \omega_1 a_1 a_2 e^{(k_1-k_2)z} \sin(\xi_1 - \xi_2),
\end{align} 
where $k_1 > k_2$ without loss of generality, and the sum harmonic vanishes for unidirectional deep water waves (the same is not true of the free surface elevation, whose prominent sum harmonics can be seen in Figure \ref{fig:Bichromatic-free-surface-constituents}). The combination of two harmonics $k_1$ and $k_2$ (or equivalently $\omega_1$ and $\omega_2$) leads to an unsteady velocity field, which can nevertheless be chosen to be exactly periodic in either space or time, and approximately periodic in the other variable. Selecting commensurable frequencies, Eulerian averaging can be taken over the period $T$ of the bichromatic wave train, showing 
\[ u_E = \frac{1}{T} \int_0^T \phi_x dt = 0 \]
for $\phi$ obtained from the 2nd order solution \eqref{eq:Second_order_bichromatic} or the 3rd order solution (Sec.\ \ref{ssec:Recovery of the third-order solution}).

\begin{figure}
\centering
\includegraphics[width=\linewidth]{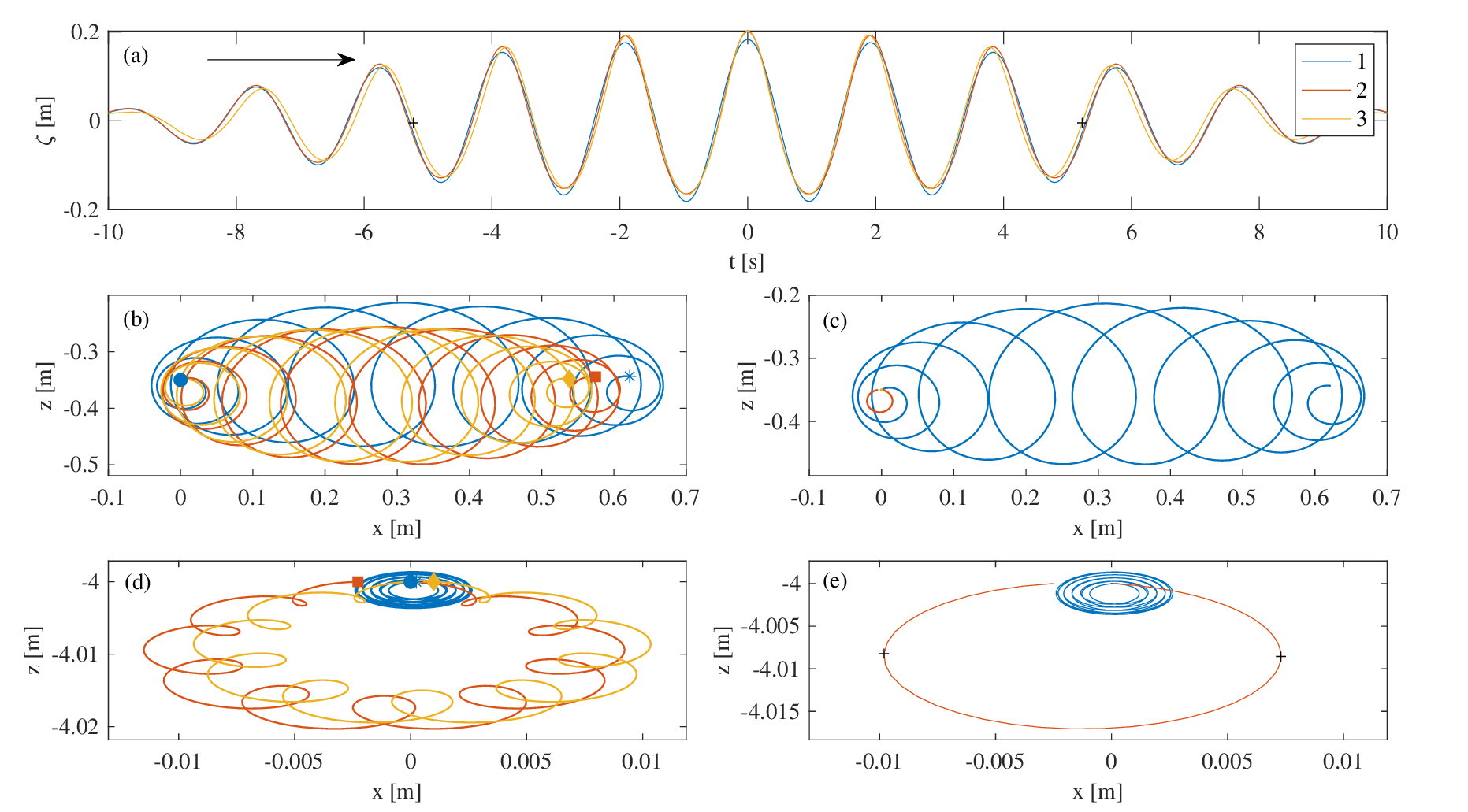} 
\caption{Time series of a bichromatic wave with $k_1 = 1.2$ and $k_2=1$ 1/m, $\epsilon_1=\epsilon_2=0.1$ at $x=0$ (a), and accompanying particle trajectories at $z_0=-0.35$ m (b)--(c) and $z_0=-4$ m (d)--(e). Blue curves denote 1st order theory, red curves 2nd order theory, and yellow curves 3rd order theory in all panels. Note that panels (c) and (e) show particle paths from 1st order theory, 2nd order contributions only (red curves) and 3rd order contributions only (yellow curves), \textit{without} the attendant lower order velocities. Markers '+' in panels (a) and (e) demarcate times between $t\approx -5.2..5.2$ s when the flow at depth is opposite the direction of wave propagation.}
\label{fig:Bichromatic-free-surface-and-pp}
\end{figure}

We show part of the free-surface time-series of one such bichromatic solution in Figure \ref{fig:Bichromatic-free-surface-and-pp}(a), where the inclusion of 2nd and 3rd order terms can be seen to have only a minor effect. This is not the case when considering the particle paths, which exhibit notable differences between linear and higher-order theories, particularly at depth. Figure \ref{fig:Bichromatic-free-surface-and-pp}(b) shows the paths of particles close to the surface with initial position $(x_0,z_0)=(0,-0.35)$ (marked by $\bullet$) and different final positions (marked with $*$, $\blacksquare$ and $\blacklozenge$ in the respective colours). Close to the surface, the particle paths at each order look qualitatively similar. The same cannot be said at depth (panel (d)), where the first-order particle paths (shown in blue) are dramatically different from the second order (red) and third order (yellow) trajectories.

Note that Figures \ref{fig:Bichromatic-free-surface-and-pp}(b),(d) are simply obtained by integrating $\mathbf{x}'(t) = \mathbf{u}(x,z,t) = \mathbf{u}^{(1)}(x,z,t)+\mathbf{u}^{(2)}(x,z,t)+\mathbf{u}^{(3)}(x,z,t),$ where $\mathbf{u}^{(i)}(x,z,t)$ denotes the velocity at a given order ($i=1,2$ or 3). The relative importance of each term can be assessed by integrating $\mathbf{x}'(t)=\mathbf{u}^{(i)}(x,z,t)$ separately for each $i$, which yields Figures \ref{fig:Bichromatic-free-surface-and-pp}(c),(e). From these it is clear that the principal contribution at the surface comes from the 1st order velocities, with the principal contribution at depth coming from 2nd order velocities. The 3rd order contributions (in yellow) are barely visible near $(x,z)=(0,-4)$ in panel (e).

\subsection{A Stokes drift approximation from second-order wave theory}

The second order solution \eqref{eq:Second_order_bichromatic} lends itself to a calculation of a modified Stokes drift in a rather straightforward manner: calculate $(u,w)$ from \eqref{eq:Second_order_bichromatic}, Taylor expand about a point $(x_0,z_0)$, keep only the constant terms in the Taylor expansion and calculate approximate trajectories $(x(t),z(t))$, subsequently substitute these trajectories into the linear terms in the Taylor expansion and integrate the resulting particle trajectory equations in time. The result of this procedure is
\begin{equation} u_S =    \underbrace{a_1^2 k_1 \omega_1 e^{2 k_1 z_0} + a_2^2 k_2 \omega_2 e^{2 k_2 z_0}}_{(\text{I})} + \underbrace{a_1^2 a_2^2 \omega_1^2 \frac{(k_1-k_2)^3}{\omega_1-\omega_2} e^{2(k_1-k_2)z_0}}_{(\text{II})} , \label{eq:Bichromatic_2nd_order_Stokes_drift} \end{equation}
which consists of the sum of the Stokes drifts of each mode separately (I) as well as a new term\footnote{The author is grateful to one of the reviewers for pointing out that, while this paper was under review, an article by \cite{liao2025mass} appeared which also derives this term, Section 1 in their Appendix C.} (denoted (II)) that arises from the difference harmonics (recall that we assume $k_1>k_2$). Just as the Stokes drift derived from linear theory includes quadratic terms, so the modified Stokes drift derived from second order theory now includes (formally) quartic terms.

\begin{figure}
\centering
\includegraphics[width=\linewidth]{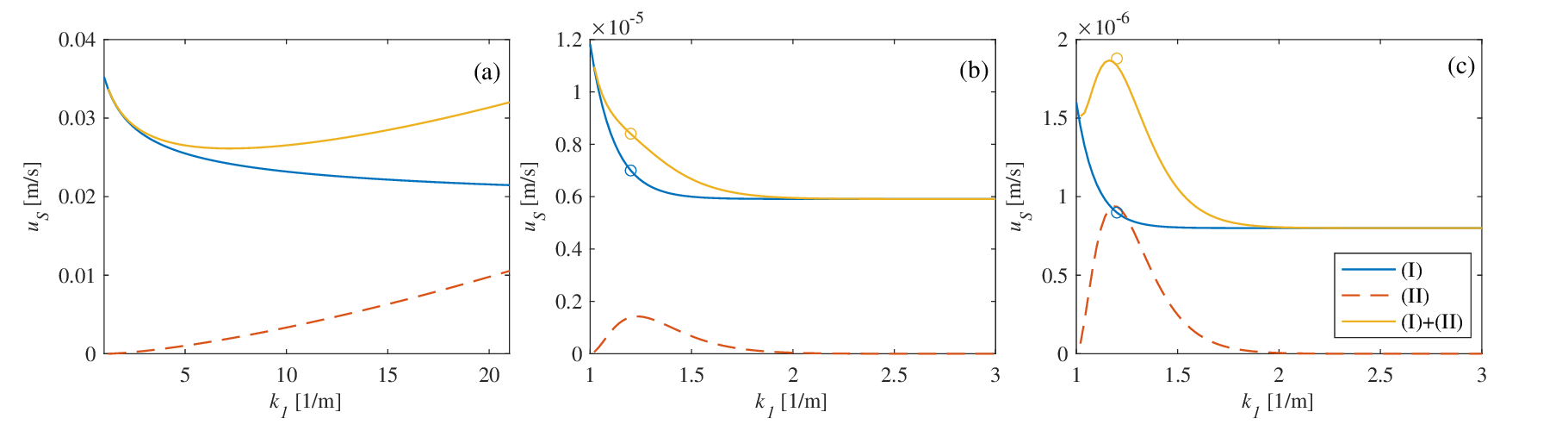} 
\caption{Illustration of the bichromatic Stokes drift approximation \eqref{eq:Bichromatic_2nd_order_Stokes_drift} for $k_2 = 1$ 1/m, $\epsilon_1=\epsilon_2=0.075$ and variable $k_1>k_2$, shown at a depth $z_0=0$ m (a), $z_0=-4$ m (b) and $z_0=-5$ m (c). Circles in panels (b) and (c) show the results of computing the drift by integration of the particle paths using the 1st order and 2nd order velocity field for $k_1=1.2$ 1/m at each depth.}
\label{fig:Bichromatic-2nd-order}
\end{figure}

While overall particle motion decreases as we descend into the fluid, there are regions where the 2nd order contribution to Stokes' drift has a sizeable effect, as we show in Figure \ref{fig:Bichromatic-2nd-order}, which plots the terms (I) and (II) of \eqref{eq:Bichromatic_2nd_order_Stokes_drift} at three depths: (a) $z_0=0$, (b) $z_0=-4$ and (c) $z_0=-5$ for a fixed wave $k_2=1$ 1/m. In connection with this, it is important to recall that the formulation \eqref{eq:Bichromatic_2nd_order_Stokes_drift} formally assumes deep water \textit{also for the difference harmonic} $k_1-k_2$, i.e.\ that $(k_1-k_2)h$ is large.

It is simplest to analyse the relative importance of the terms of \eqref{eq:Bichromatic_2nd_order_Stokes_drift} by setting $k_1=1+\Delta$ and varying $\Delta>0.$ We note that the contribution (II) at the top of the fluid domain $z_0=0$ grows monotonically with $\Delta$, as can be observed in Figure \ref{fig:Bichromatic-2nd-order}(a) (note that the steepness of both modes is fixed at $\epsilon=0.075$). By contrast, for $z_0<0$ the quartic term (II) has a single maximum and subsequently decays, as shown in Figure \ref{fig:Bichromatic-2nd-order}(b)--(c). As we descend into the fluid, this maximum moves towards smaller values of $k_1.$ At certain depths, and for certain bichromatic waves, the second-order contribution (II) dominates the 1st order contribution (I), as shown around $k_1=1.2$ in panel (c). These contributions at depth remain several orders of magnitude smaller than the surface Stokes drift, although their significance (or lack or significance) should be assessed based on the time-scales of interest, and over the entire water column (see Section \ref{ssec:Parametric spectra} below).

\subsection{Lagrangian drift for higher-order bichromatic waves}

In attempting to compare the formula \eqref{eq:Bichromatic_2nd_order_Stokes_drift} with other methods of obtaining the wave-induced drift from wave theory, the principal difficulty we encounter -- for the first time with bichromatic waves -- is connected to the unsteadiness of the flow field and the lack of a unique solution towards which different expansions are known to converge. Unlike the monochromatic waves of Section \ref{sec:Monochromatic waves}, we cannot simply compare waves of the same length and height, which differ geometrically only in the curvature of the profile between crest and trough. An additional complication arises at third order with the appearance of an asymmetric frequency correction (for $k_1>k_2$) first found by \cite{Longuet-Higgins1962d}
\begin{align} \label{eq:Omega-nonlin-bichrom-short-wave-1}
\Omega_1 &= \omega_1 \left[ 1+ \frac{1}{2} \epsilon_1^2 + \epsilon_2^2 \left( \frac{k_1}{k_2} \right)^{1/2} \right],\\ \label{eq:Omega-nonlin-bichrom-long-wave-2}
\Omega_2 &= \omega_2 \left[ 1+ \frac{1}{2} \epsilon_2^2 + \epsilon_1^2 \left( \frac{k_2}{k_1} \right)^{3/2} \right].
\end{align}
This means that bichromatic waves at third-order (and 5th, 7th, and so on) will be out of phase with their lower order counterparts, as seen in Figure \ref{fig:Bichromatic-free-surface-constituents}.

This is compounded by a paucity of results (theoretical or experimental) on drift associated with bichromatic waves, making comparisons difficult. However, we can employ the methodology tested for monochromatic waves in Section \ref{sec:Monochromatic waves} to obtain results for bichromatic waves at the surface and below the trough level, and compare these to the approximate Stokes drift derived in \eqref{eq:Bichromatic_2nd_order_Stokes_drift}. In this connection, it is important to emphasise that the Stokes drift velocity $u_S$ is still defined as the difference between Eulerian and Lagrangian mean velocities. In dealing with periodic, bichromatic wave groups, our Eulerian mean velocity (with the mean taken over the period of the group) vanishes exactly as for monochromatic waves. The terms of $u_S$ in \eqref{eq:Bichromatic_2nd_order_Stokes_drift} are therefore an \textit{approximation} to the Lagrangian mean velocity which takes difference harmonics into account.

\subsubsection{Lagrangian drift with depth}

\begin{figure}
\centering
\includegraphics[width=\linewidth]{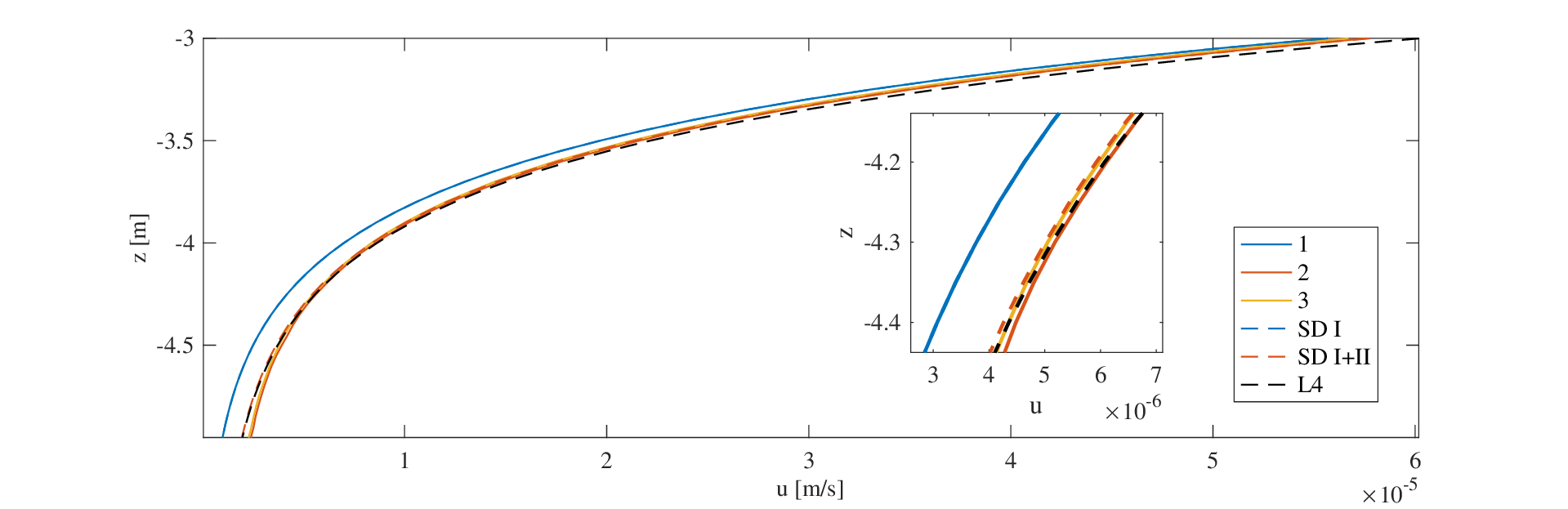}
\caption{Comparison of horizontal drift velocities with depth for a bichromatic wave train with $k_1=1.2$ and $k_2=1$ 1/m with $\epsilon_1=\epsilon_2=0.075,$ as in Figure \ref{fig:Bichromatic-free-surface-constituents}  }
\label{fig:Bichromatic-Stokes-drift-at-depth}
\end{figure}

Below the surface, exactly as shown for monochromatic waves, it is possible to use the particle trajectories obtained by integrating the Eulerian velocity field to obtain the Lagrangian drift. This is put to the test in Figure \ref{fig:Bichromatic-Stokes-drift-at-depth}, where we select a bichromatic wave train with significant subharmonic drift as per \eqref{eq:Bichromatic_2nd_order_Stokes_drift}, namely $k_1=1.2$ and $k_2=1$ 1/m. This corresponds to the circles shown in Figure \ref{fig:Bichromatic-2nd-order}. 

Figure \ref{fig:Bichromatic-Stokes-drift-at-depth} compares the Stokes drift formulations (I) and (I)+(II) of \eqref{eq:Bichromatic_2nd_order_Stokes_drift} with the 4th order Lagrangian drift of \cite{blaser2025increased}, and the Lagrangian drift obtained from integration of the particle trajectory mapping \eqref{eq:particle-trajectory-mapping} with 1st, 2nd and 3rd order velocity fields. The 2nd order horizontal velocity can immediately be read off from \eqref{eq:Second_order_bichromatic}, while the 3rd order velocity, for which a simple algebraic expression has not been found, is obtained from the Hamiltonian expansion of Section \ref{sec:WW problem and Hamiltonian formulation}, eq.\ \eqref{eq:phi-to-third-order}. 

The Lagrangian drift obtained from integrating the 1st order theory overlaps completely with the classical Stokes drift (SD I). On the other hand, there is very good agreement between 2nd and 3rd order theories, the improved Stokes drift formulation (SD I+II) and the fourth order Lagrangian theory (L4). This points to the fact that the approximation \eqref{eq:Bichromatic_2nd_order_Stokes_drift} captures important features of the bichromatic Lagrangian drift. The same comparison for a pair of more widely separated wavenumbers (such as $k_1=2$ and $k_2=1$ 1/m, not shown) would demonstrate that all six curves essentially coincide, as might be predicted from Figure \ref{fig:Bichromatic-2nd-order} (noting that there are no notable flows induced by the difference harmonic in this case).

\subsubsection{Lagrangian drift at the surface}

To obtain drift at the surface we integrate the equation for the horizontal particle position $x'(t)=u(x,\zeta,t),$ where the velocity field is given by Taylor expansion about $z=0$ to a given order, exactly as in Section \ref{ssec:Mono-Lag-drift-at-surface} for monochromatic waves. In addition to the velocity fields to third-order, we also require the second-order bichromatic free surface, which is given in \eqref{eq:zeta2}.

\begin{figure}[h]
\centering
\includegraphics[width=\linewidth]{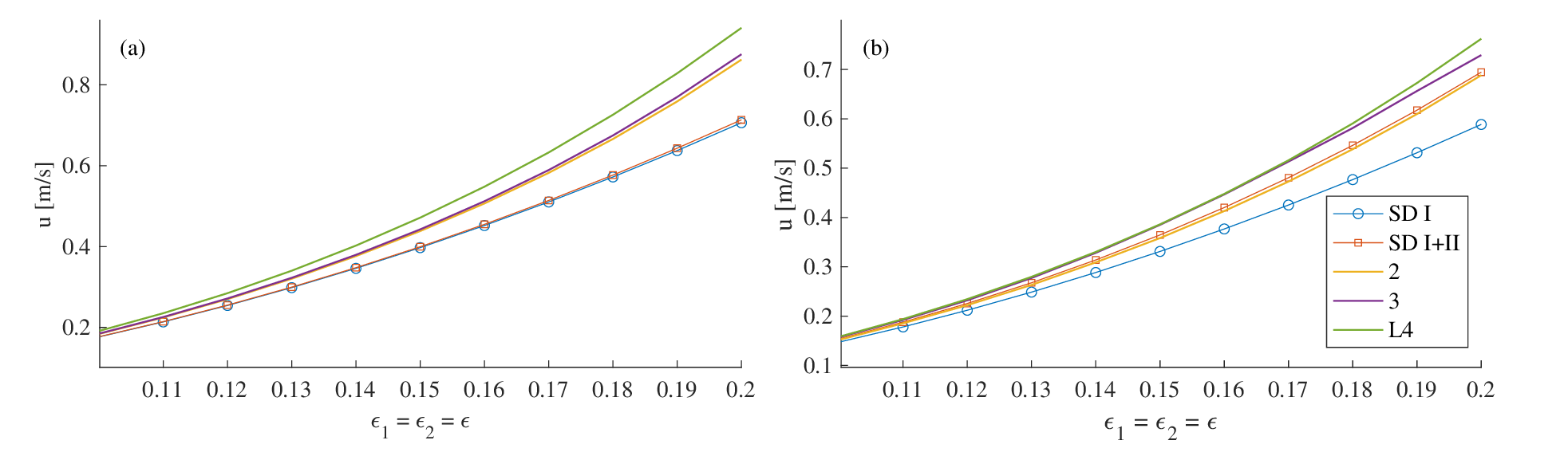} 
\caption{Comparison of surface drift formulations for a bichromatic wave train with (a) $\omega_1=1.25, \omega_2=1$ rad/s and (b) $\omega_1=2, \omega_2=1$ rad/s and identical (linear) steepness $\epsilon_1=\epsilon_2$. SD I and SD I+II denote the respective terms in \eqref{eq:Bichromatic_2nd_order_Stokes_drift}. These are compared with the 2nd order and 3rd order bichromatic solution, as well as the 4th order Lagrangian drift (L4) of \cite{blaser2025increased}.}
\label{fig:Bichromatic-surface-Stokes-drift}
\end{figure}

Figure \ref{fig:Bichromatic-surface-Stokes-drift} compares the Lagrangian surface drift of 2nd and 3rd order theory with the classical Stokes drift (SD I) and the modified Stokes drift (SD I+II) evaluated at $z_0=0.$  In panel (a), where the two waves are close in frequency, we expect the contributions of the terms (II) to be negligible. Instead, the Lagrangian drift at the surface is influenced by sum-harmonics appearing in the 2nd and 3rd order theory, and which are not accounted for by the approximation \eqref{eq:Bichromatic_2nd_order_Stokes_drift}.

When the constituent harmonics are more widely separated, such as in the case of $\omega_1=2$ and $\omega_2=1$ rad/s shown in panel (b), the difference harmonic contributions of (II) are more prominent. In both cases the 2nd and 3rd order theories give an enhanced drift, which is consistent with the results obtained for monochromatic waves and shown in Figure \ref{fig:Surface-Stokes-drift-monochromatic}.

Figure \ref{fig:Bichromatic-surface-Stokes-drift} also compares with the 4th order Lagrangian drift obtained from the 3rd order bichromatic solution in Lagrangian variables obtained in Appendix B of \cite{blaser2025increased}, which generalises Pierson's \citeyearpar{Pierson1961} classical 2nd order result. Unfortunately, bichromatic waves in Eulerian and Lagrangian coordinates are not the same; rather, the wave profiles are geometrically distinct, with a phase-shift that occurs between successive crests of both solutions, as already noted by \cite{fouques2008comparing}\footnote{This phase shift is not due to differences in dispersion relation, as might be hypothesised from Pierson \citeyearpar{Pierson1961}, since Blaser et al \citeyearpar{blaser2025increased} recover the Eulerian dispersion correction \eqref{eq:Omega-nonlin-bichrom-short-wave-1}--\eqref{eq:Omega-nonlin-bichrom-long-wave-2} at 3rd order.}. Whether it is possible to adapt the Lagrangian perturbation expansion to match the Eulerian solution (or vice versa) remains an open question. For well-separated frequencies (panel (b)) the Lagrangian drift obtained from our Eulerian particle trajectory mapping \eqref{eq:particle-trajectory-map-surface} and the Lagrangian solution of \cite{blaser2025increased} remain remarkably close, while for two nearby frequencies (panel (a)) the notable difference grows with wave slope.

\subsection{Difference harmonics and particle drift}
\label{ssec:Bichromatic difference harmonics and drift}

In general, we see the importance of leading order effects near the surface of bichromatic waves, while second order effects become more prominent at depth. This is no surprise, and follows from the structure of the difference harmonic terms themselves. As a prelude to our work on wave groups it is expedient to investigate their role in bichromatic seas. 

We first consider the velocity field which drives the particle trajectory mapping, and which can be calculated from the second-order \eqref{eq:Second_order_bichromatic} or third-order potential \eqref{eq:phi-to-third-order}. In fact, Figure \ref{fig:Bichromatic-free-surface-and-pp} makes clear that the 3rd order velocity field -- though difficult to calculate -- makes only a small contribution. In large part this is due to the 3rd order dispersion correction \eqref{eq:Omega-nonlin-bichrom-short-wave-1}--\eqref{eq:Omega-nonlin-bichrom-long-wave-2}, as can be seen by comparing panels (b) and (d) with panels (c) and (e) in Figure \ref{fig:Bichromatic-free-surface-and-pp}. These dispersion corrections do not affect the instantaneous velocity field, only its time evolution.

\begin{figure}
\centering
\includegraphics[width=\linewidth]{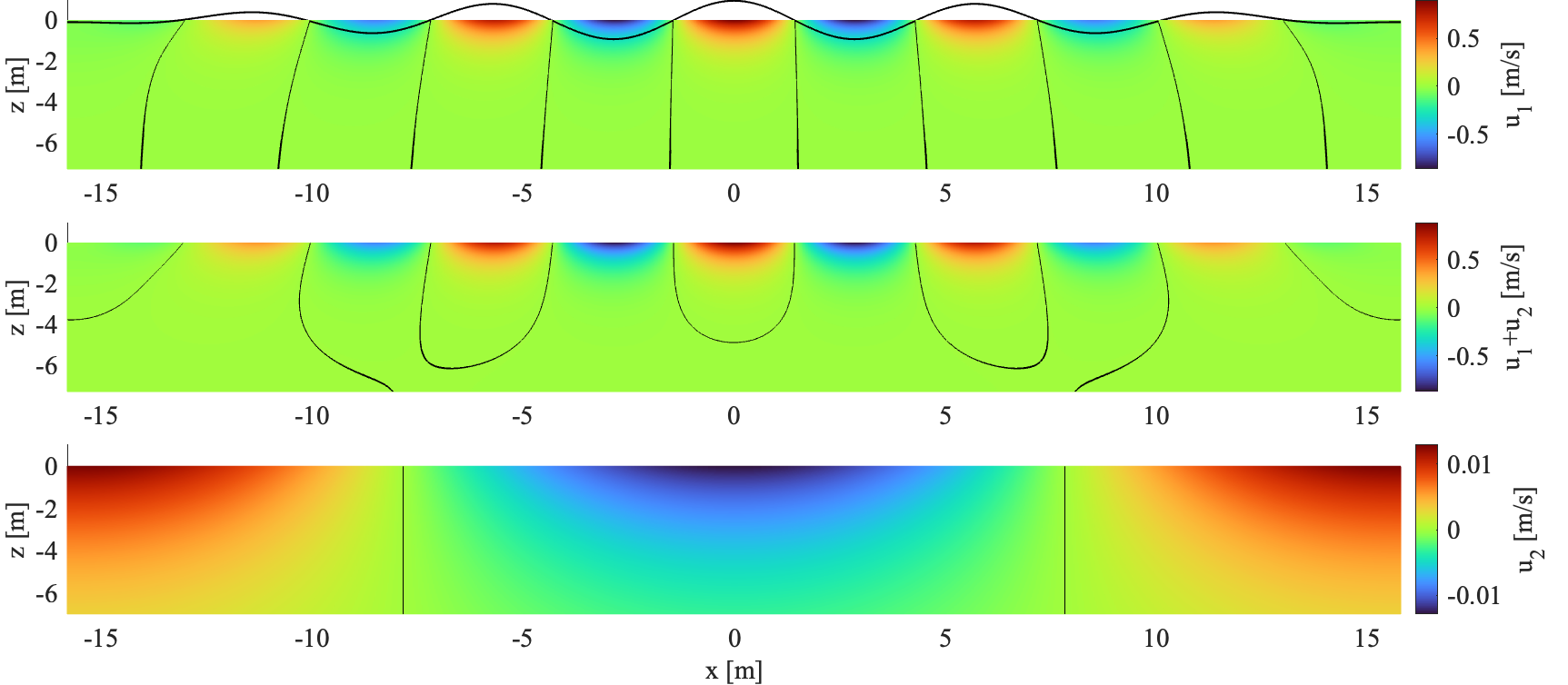}
\caption{Horizontal velocity for a bichromatic wave train with $k_1=1.2$ and $k_2 = 1$ 1/m, shown indicatively in the top panel. (Top) First order velocity only. (Middle) Sum of first and second order horizontal velocities. (Bottom) Second order horizontal velocity only. Black curves are contours of vanishing horizontal velocity.}
\label{fig:Bichromatic-horizontal-velocity-contours}
\end{figure}

As an example we consider the waves studied in Figure \ref{fig:Bichromatic-2nd-order} with $k_1=1.2$ and $k_2=1$ 1/m, and focus only on horizontal velocities.
When only the first order contributions $u_1$ are retained we find the familiar pattern of forward velocity below the crests and backward velocity below the troughs, with an exponential fall-off in depth as depicted in the top panel of Figure \ref{fig:Bichromatic-horizontal-velocity-contours}.
If instead we focus only on the second-order contributions $u_2$, the situation looks quite different. The exponential decay in velocities is slower because $\exp((k_1-k_2)z)$ is larger than either $\exp(k_1 z)$ or $\exp(k_2 z)$ for $k_1 \sim k_2,$ and we find a negative horizontal velocity under the centre of the bichromatic wave train, as shown in the bottom panel of Figure \ref{fig:Bichromatic-horizontal-velocity-contours}. The full second-order picture is a sum of the first and second order terms, shown in the middle panel of the same figure, and it is clear that the first order dominates near the surface while the second-order difference harmonic dominates at depth. This can be observed clearly by considering the zero contour lines of the horizontal velocity, shown in black for each case.

The generic situation for particle paths in a bichromatic wave train is thus as follows: close to the surface the first order contributions dominate, giving the curlicue particle paths observed in  Figure \ref{fig:Bichromatic-free-surface-and-pp}(b). The effect of difference harmonics is to retard slightly the forward motion of the particles, due to the slow moving (in deep water the group velocity is half the phase velocity) region of backward (negative) velocity under the highest crests observed in Figure \ref{fig:Bichromatic-free-surface-and-pp}(e) and Figure \ref{fig:Bichromatic-horizontal-velocity-contours}.

Deeper into the fluid column the effect of the first order free-modes becomes negligible. The fluid motion is dominated by the difference harmonic term $k_1-k_2$, and the particle paths behave essentially as those of a monochromatic wave with that wavenumber. This is clearly illustrated in Figure \ref{fig:Bichromatic-free-surface-and-pp}(e), where we see that the modes $k_1$ and $k_2$ which make up the first-order theory are only a small perturbation on the displacement due to the difference harmonic term.

It is worth pointing out the existence of a region beneath the bichromatic wave train where the particle displacement is largely against the direction of wave propagation, as seen between $t\approx[-5.2,5.2]$ s in Figure \ref{fig:Bichromatic-free-surface-and-pp}(e) (see markers '+'). This should not be surprising, since it simply reflects the effective motion of the difference harmonic term as shown in panel (d). Despite the existence of this `return flow' -- which we shall encounter again in our discussion of wave groups -- the average particle drift over the period of the bichromatic group is in the direction of wave propagation at every depth in the fluid, as shown in Figure \ref{fig:Bichromatic-2nd-order} and Figure \ref{fig:Bichromatic-Stokes-drift-at-depth} below. Because the solution is periodic, over many periods the net effect of difference harmonics is to enhance forward drift, as shown in Figure \ref{fig:Bichromatic-Stokes-drift-at-depth}.

\section{Multichromatic waves}
\label{sec:Multichromatic waves}

When more than two Fourier modes are necessary to describe the linear problem, we call such a configuration multichromatic. The second order contains sums and differences of any two modes, while the third order contains sums and differences of any three. This means that significant algebra can be involved in calculating higher-order solutions. 

\subsection{Focused linear wave groups}

In treating waves with an increasing number of linear harmonic constituents we also have a plethora of parameters that can be tuned. One case of interest, from both an experimental and a theoretical standpoint is that of a (focused) wave group. Linear focusing is quite straightforward: if our free surface is a superposition of sinusoids as per \eqref{eq:lin-solution}, then writing 
\[ \zeta_1 = \sum_i a_i \cos(k_i(x-x_0) - \omega_i(t-t_0)) \]
readily gives a wave train that focuses -- i.e.\ all components come into phase -- at $(x_0,t_0)$. Pulse-like wave groups can be created by suitable tuning of the amplitude spectrum $a_i.$ However, for any finite number of Fourier modes the group so created will never be completely localised; depending on the choice of frequencies $\omega_i$ or wavenumbers $k_i$ it may or may not be periodic in time or space, and the extent to which it appears to have a well-formed envelope depends on the choice of harmonics and amplitude spectrum.

\begin{figure}[h]
\centering
\includegraphics[width=\linewidth]{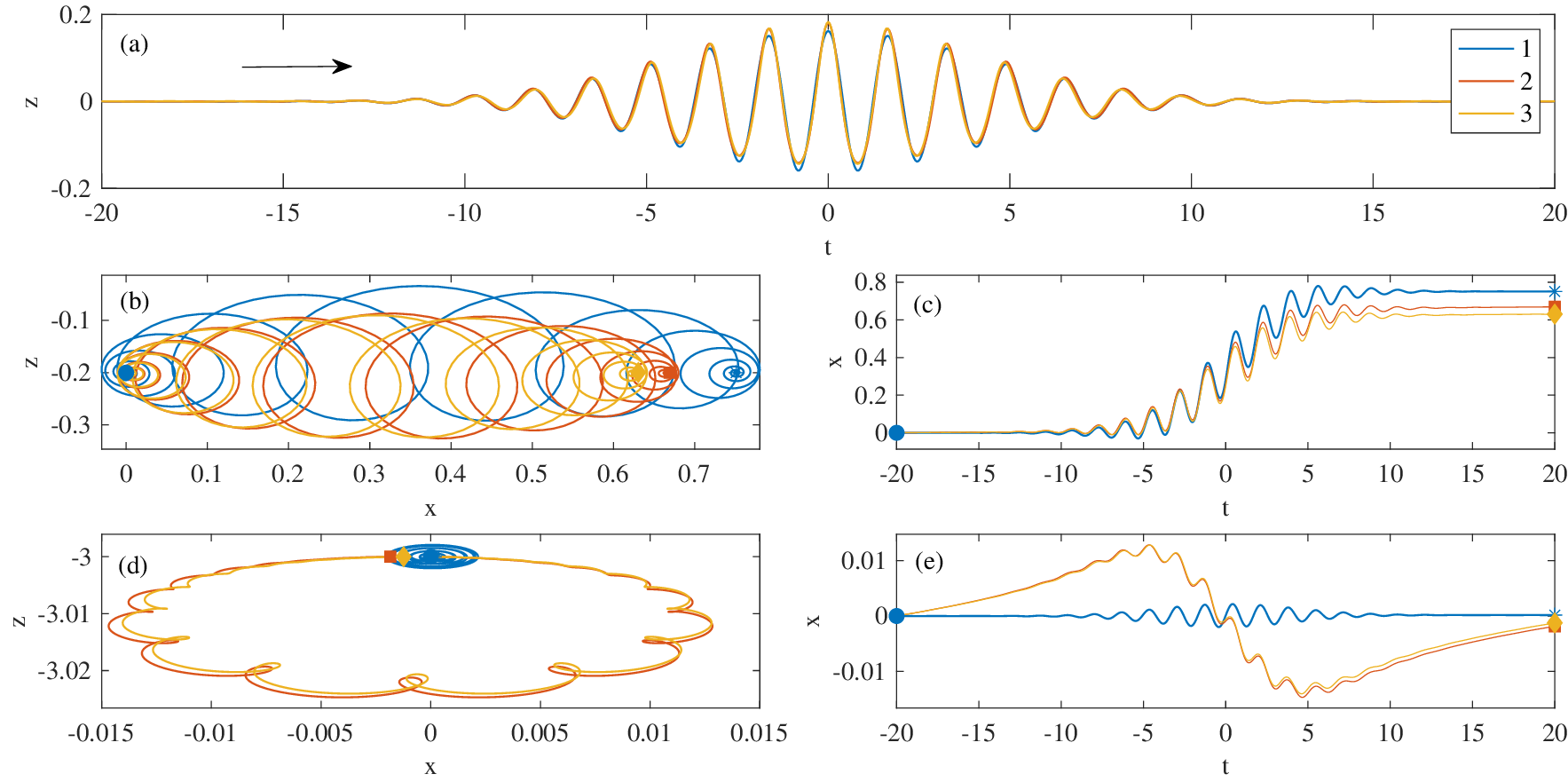}
\caption{A crest-focused wave group with $k_p=1.5, \sigma=0.18$ and 10 Fourier modes, with $S=0.24.$ Panel (a) shows the free surface in 1st, 2nd and 3rd order theory. Panels (b) and (d) show the respective particle trajectories with initial position $(x_0,z_0)=(0,-0.2)$ and $(0,-3),$ beginning at time $t=-20$ prior to the passage of the group. Panels (c) and (e) show the horizontal particle position of panels (b) and (d) with time $t,$ illustrating the negative drift at depth also seen in Figure \ref{fig:Bichromatic-free-surface-and-pp}.}
\label{fig:Focused-wave-group-surface-and-pp}
\end{figure}

One possible choice is a Gaussian amplitude spectrum of the form 
\begin{equation} \label{eq:Gaussian-amplitude-spectrum}
a_i = A \exp\left( -\frac{(k_i-k_p)^2}{2 \sigma^2}\right).
\end{equation}
An example of a wave train with such an amplitude spectrum is shown in Figure \ref{fig:Focused-wave-group-surface-and-pp}(a), where we have chosen $10$ equidistant modes between $k_1=1$ and $k_{10}=2$ 1/m, $k_p=1.5$, $A=0.04$ and $\sigma=0.18.$ The focusing location is chosen to be $(x_0,t_0)=(0,0)$, and the focus occurs at a crest.
Following \cite{blaser2025increased} we define 
\[ S = \sum_{n=1}^N a_n k_n \]
as the maximum slope for the linear focused wave, and use this as a measure of nonlinearity going forward.

The particle paths in Figure \ref{fig:Focused-wave-group-surface-and-pp}(b),(d) are in many ways similar to those in the simpler, bichromatic wave group shown in Figure \ref{fig:Bichromatic-free-surface-and-pp}. Towards the top of the fluid column, 1st order theory dominates, while at depth the effect of the in-phase subharmonics associated with higher-order theories clearly accounts for the bulk of the particle motion. In order to further illustrate this point, panels (c) and (e) show only the horizontal component of the particle position as a function of time $t$, as the wave group passes over the initial position. Towards the surface (c) there is nearly no motion until $t=-10$ s, while at depth we observe a steady forward drift in the positive $x$-direction. Under the centre of the group, as expected, we see a strong forward drift towards the surface (panel c) and backward (return) flow at depth (panel e).

\begin{figure}
\centering
\includegraphics[width=\linewidth]{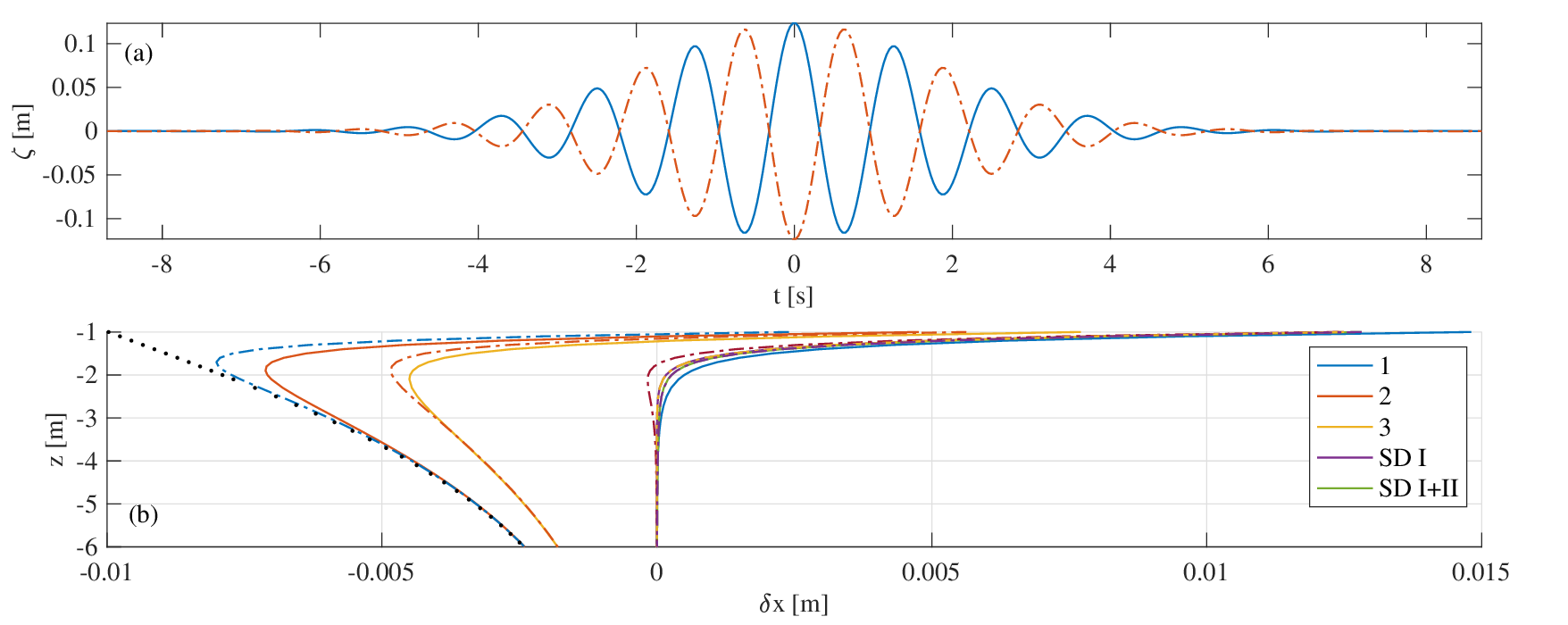} 
\caption{Indicative horizontal drift $\delta x$ beneath a crest/trough focused wave train. Panel (a) shows a crest focused (blue) and trough focused (red, dash-dotted) Gaussian wave group with $k_p=2.5$ 1/m, $\sigma=0.55$ and 10 equispaced modes between $k_1=1$ and $k_{10}=4$ 1/m, with steepness $S=0.3.$ Panel (b) shows the horizontal (Lagrangian) displacement $\delta x$ of a particle as the wave group passes over it (from $t=-8.7$ to $t=8.7$ s, as shown in panel (a)). Solid lines show the drift for the crest focused wave, dash-dotted lines the corresponding trough-focused values. Black dots refer to the second-order net displacement of the return flow calculated from (2.10) of \cite{higgins2020lagrangian}.}
\label{fig:Horizontal-drift-focused-waves}
\end{figure}

This effect can be seen even more clearly in Figure \ref{fig:Horizontal-drift-focused-waves}, which illustrates (for a single particle) the depth dependence of the horizontal motion. The (leading order) free surface of a focused Gaussian wave train with $k_p=2.5$ 1/m is shown in panel (a), in both the crest focused (blue, solid curve) and trough focused (red, dash-dotted curve) case. Panel (b) shows the horizontal displacement $\delta x$ of a particle as the wave group passes it. \cite{higgins2020lagrangian} suggest that the net (long-time) displacement of the return flow to second order can be calculated by integrating the second-order velocity $u_2$ in time (see their eq.\ (2.10)), and we see (panel (b), black dotted curve) that this indeed matches our second order displacement (shown in red) quite well at depth.

Interestingly, there is a marked difference in the return flow between crest and trough focused cases, with even 1st order theory showing some return flow for trough focused waves (dash-dotted curves). Such (localised) return flows are not captured by Stokes drift approximations, whether based simply on the linear theory or incorporating second order effects via the extension of \eqref{eq:Bichromatic_2nd_order_Stokes_drift}
\begin{equation} \label{eq:multi-mode-2nd-order-Stokes-drift}
u_S = \underbrace{\sum_j a_j^2 \omega_j k_j e^{2k_j z_0}}_{\text{(I)}} + \underbrace{\sum_{k_i>k_j} \frac{\omega_i ^2 a_i^2 a_j^2 (k_i-k_j)^3}{\omega_i-\omega_j} e^{2(k_i-k_j)z_0}}_{\text{(II)}}.
\end{equation}
Indeed, the effects shown in Figure \ref{fig:Horizontal-drift-focused-waves} are highly phase-dependent, and rely on suitable choice of position in the spatio-temporal evolution of the group, while the approximate formulae \eqref{eq:multi-mode-2nd-order-Stokes-drift} can only capture an averaged drift based on the Fourier amplitudes, but independent of the phases. This contrasts strongly with the theoretical result of \cite{higgins2020lagrangian} for a spatially localised packet. 

\begin{figure}
\centering
\includegraphics[width=\linewidth]{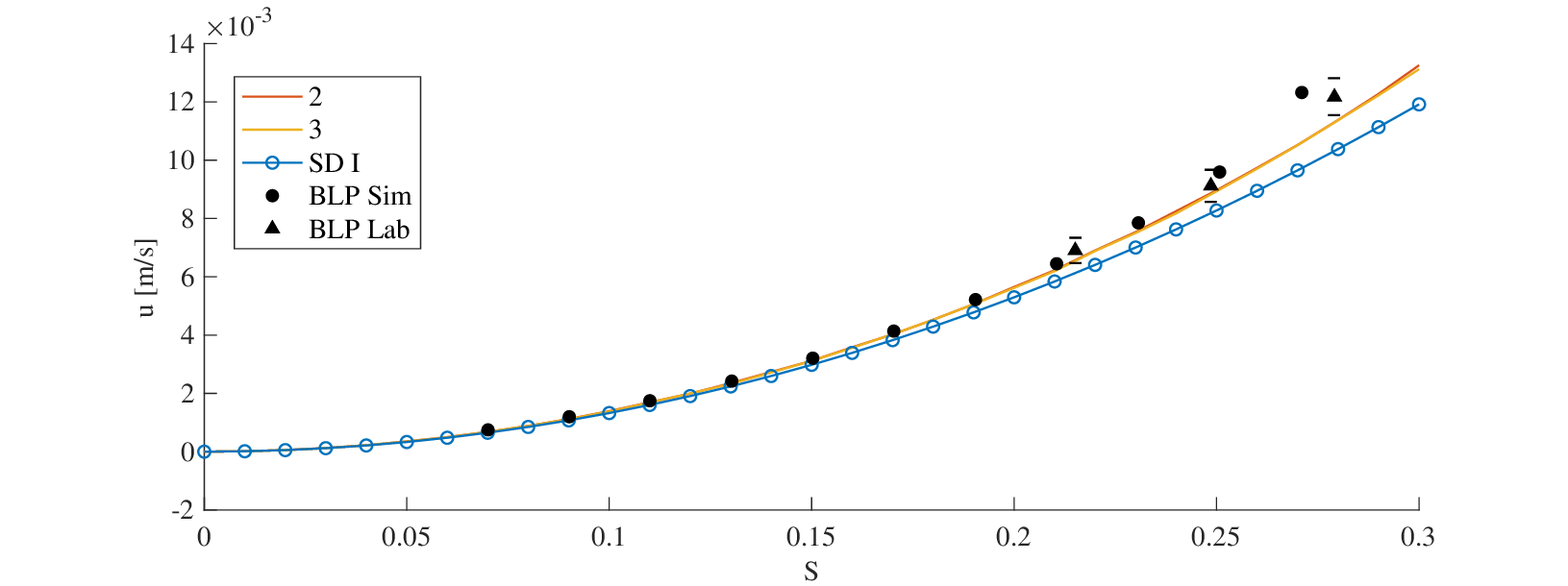}
\caption{Comparison of surface Lagrangian drift beneath a wave group using 2nd and 3rd order theories, as in Figures \ref{fig:Surface-Stokes-drift-monochromatic} \& \ref{fig:Bichromatic-surface-Stokes-drift}, compared with the leading order approximation to the Stokes drift (SD I). The dimensionless bandwidth parameter $\Delta=0.77$ and steepness $S$ is varied from 0 to 0.3 to compare with Figure 3(a) of \cite{blaser2025increased}. Circles ($\bullet$) denote their fully nonlinear numerical simulations (with $\Delta=0.8$) and triangles ($\blacktriangle$) denote the mean experimental result.}
\label{fig:Surface-Drift-BLP-comparison}
\end{figure}

As we have done for monochromatic and bichromatic waves, we can apply 2nd and 3rd order, deep-water theory directly to calculate the Lagrangian surface drift of focused wave groups, using the particle trajectory mapping at the surface. The results of such a computation are reported in Figure \ref{fig:Surface-Drift-BLP-comparison}, which is the counterpart of Figures \ref{fig:Surface-Stokes-drift-monochromatic} and \ref{fig:Bichromatic-surface-Stokes-drift}. In this case, we have initialised our wave group with a central frequency $\omega_c=2 \pi,$ and using $N=12$ modes have defined the amplitudes $a_n=S(Nk_n)^{-1},$ where the frequencies are chosen via $\omega_n=\omega_c(1+\Delta(n-N/2)/N)$ for bandwidth $\Delta=0.77,$ and $k_n$ are the wavenumbers obtained from the linear dispersion relation. This is precisely the formulation used by \cite{blaser2025increased}, and allows us to compare our results with their experimental and numerical values.

Following the trends for monochromatic and bichromatic waves, we note that the lowest order Stokes drift evaluated at $z_0=0$ (SD I) and which treats the Stokes drift as a sum of the drifts of each individual mode \eqref{eq:lin-monoch-Stokes-drift} gives the smallest value. The 2nd and 3rd order formulations show good agreement with both laboratory experiments ($\blacktriangle$) and fully nonlinear simulations ($\bullet$) up to rather high steepness $S$, even though our formulation neglects amplitude evolution which is included in the numerical results (see also the discussion in Appendix \ref{app:Amplitude evolution}).

A further issue which we encounter in these calculations is the lack of joint spatio-temporal localisation of our wave groups. This fact obscures the scale over which a ``mean" quantity such as $u_E$ should be calculated, and implies that we cannot always numerically enforce $u_E \equiv 0$ by suitable construction of the group (see the discussion in Section \ref{sec:Bichromatic waves}). Thus, while it is possible to compare the Lagrangian drift obtained from integration of particle paths with the Lagrangian drift obtained computationally and experimentally (as in Figure \ref{fig:Surface-Drift-BLP-comparison}), the difference between Lagrangian drift and (a priori unknown) Stokes drift is an Eulerian mean flow which is sensitive to the domain over which the averaging is taken.

\subsection{Random multichromatic waves}

Random waves are the closest mathematical idealisation to the waves we might typically find in the ocean: the phases of one Fourier component are not correlated with the next, and the resulting pattern -- while occasionally exhibiting a distinct group -- is highly irregular. Random phases at lowest order also means that sum and difference phases at higher orders are random, due to a neglect of slow phase-coherence effects associated with nonlinear coupling \citep{Andrade2023}. In such cases, the averaging necessary to unambiguously define an Eulerian mean velocity is not achievable. At best we can average over a suitably large number of individual waves, in the expectation that the Eulerian mean velocity over such a large time will be small.

Where focused wave groups exhibit localised features, such as a clear division between flows in the direction of propagation at the surface and return flows at depth, --  which have been well described by using wave-envelope formulations by van den Bremer and coworkers \citep{vandenBremer2016,vandenBremer2018,VandenBremer2019,calvert2019laboratory} -- the lack of clear ``groupiness" in random seas means that we should expect these localised effects to be averaged out. The advantage is that we can test whether (on average) the Stokes drift formulation \eqref{eq:multi-mode-2nd-order-Stokes-drift} is suitable, since it clearly cannot capture spatio-temporally localised flows around a group itself (all terms are positive).

\begin{figure}[h]
\centering
\includegraphics[width=\linewidth]{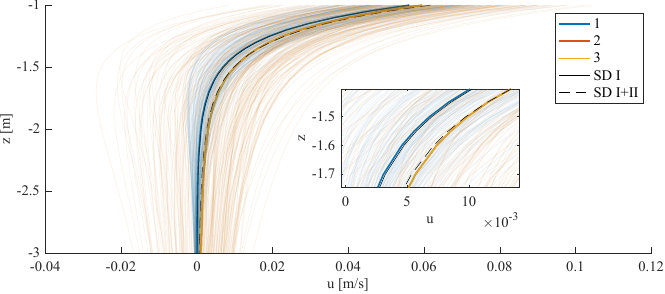}
\caption{Drift velocity with depth below a random wave field, calculated from the average of 200 realisations (shown as lighter curves) of an irregular sea-surface with random phases. The horizontal Lagrangian drift velocities are calculated from the velocity fields as in \ref{fig:Bichromatic-Stokes-drift-at-depth} using 1st order, 2nd order, and 3rd order theory, and compared with the Stokes drift approximations of \eqref{eq:multi-mode-2nd-order-Stokes-drift} (SD I) and (SD I+II).}
\label{fig:Random-waves-depth-dependent-drift}
\end{figure}

Figure \ref{fig:Random-waves-depth-dependent-drift} shows the results of computing the wave-induced particle displacement for 200 realisations of a random sea surface. A Gaussian amplitude spectrum \eqref{eq:Gaussian-amplitude-spectrum} with $A=0.05$ and the parameters $k_p=2.5$ 1/m and $\sigma=0.7$ are used, but rather than assigning the phases so as to achieve linear focusing at a specific location, these are randomly generated for each realisation. Blue curves denote first order theory, red curves second order, and yellow curves third order, with the prominent, darker curves representing the average over all realisations for a given order. In each case the integration is for approximately 80 peak periods, which is long enough to clearly distinguish the particle drift.

With a peak wavenumber $k_p=2.5$ 1/m, most of the motion due to the free-wave constituents is filtered out at a depth of $z=1.25$ m. The linear theory gives a very narrow range of displacements, with the realisations in red barely showing deviation from the average, and agreeing well with the linear Stokes drift (SD I). Second and third order displacements are quite large in some realisations, depending on the phase relationships involved. However, as expected, there is no systematic return flow at any depth for random waves, because such (spatially and temporally) localised flows are averaged out to yield the net forward drift which we observe in Figure \ref{fig:Random-waves-depth-dependent-drift}. Moreover, all theories approach zero from above with depth, exhibiting the same ordering we have hitherto observed: linear theory predicting the smallest drift followed by second and third order. The correction (II) in \eqref{eq:multi-mode-2nd-order-Stokes-drift} is found to give a useful improvement over using only the classical Stokes drift term (I).

\subsection{Parametric wave spectra}
\label{ssec:Parametric spectra}

The comparisons of the previous sections, based on explicit computation at different orders, support the inclusion of quartic difference harmonic terms in calculations of the Stokes drift throughout the water column. In the absence of information about Eulerian currents in the open ocean, the Stokes drift is the constituent of the Lagrangian drift which can be estimated directly from the wave field. As we have seen, the difference harmonic constituents provide a useful approximation to this Stokes drift.  For many applications in ocean science and engineering it is advantageous to present such a result in terms of energy rather than amplitude spectra, and to explore its consequences for some common parametric spectral shapes.

For a continuous wavenumber spectrum $E(k)$, we make the identification with the modal amplitudes 
${a_n^2}/{2} =: E_n = E(k) dk$
where the uniform grid spacing $dk$ is assumed to tend to zero. With this assumption, it is straightforward to rewrite the discrete multi-chromatic Stokes drift \eqref{eq:multi-mode-2nd-order-Stokes-drift} with second order contributions as
\begin{align} \label{eq:Spectral-SD-I-and-II}
u_S = 2 \int_0^\infty E(k) \omega(k) k e^{2kz} dk + 4 \int_0^\infty \int_{k'}^\infty \frac{\omega^2(k) E(k) E(k') (k-k')^3}{\omega(k)-\omega(k')} e^{2(k-k')z} dk dk',
\end{align}
where the first term is identical to that found by Kenyon \citeyearpar{Kenyon1969} and the second term comes from the inclusion of second-order terms in deep water.

Numerous critical discussions have focused on the key role of the high-frequency cut-off on the Stokes drift near the surface, including  \cite{breivik2014approximate} and more recent work by \cite{Clarke2018} and \cite{Lenain2020}. As Clarke and van Gorder \citeyearpar{Clarke2018} point out, if the Stokes drift is rewritten in terms of frequency $u_S \sim \int E(\omega) \omega^3 d \omega,$ and the high-frequency tail of the spectrum is such that $E(\omega)\sim \omega^{-n}$, then we have a divergent expression for $n\leq 4$. Their proposed resolution, introducing a wave breaking frequency as an upper limit for spectral Stokes drift calculations was found to be reasonable for the subsurface Stokes drift by \cite{Lenain2020}, but for the calculations below we will employ the maximum wavenumber $k_M$ suggested by \cite{Lenain2017} and used by \cite{Lenain2020}.

\begin{figure}[h]
\centering
\includegraphics[width=\linewidth]{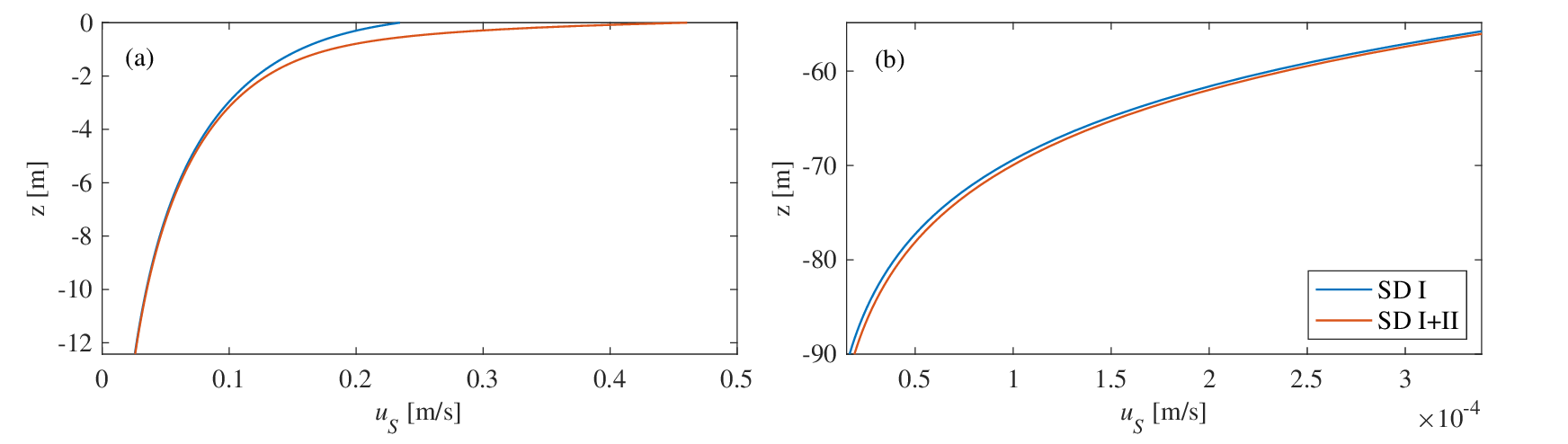}
\caption{Comparison of Stokes drift formulations for a Phillips spectrum with $T_p=10$ s. Panel (a) compares the Stokes drift near the surface up to the e-folding depth of the peak wavenumber ($k_p=0.04$ 1/m). Panel (b) compares the Stokes drift formulations at depths below one peak wavelength.}
\label{fig:Phillips-SD-Comparison}
\end{figure}

As a first comparison, we consider the Phillips spectrum for the equilibrium range of wind-generated waves, given in terms of radian frequency as 
\begin{equation}
E(\omega) = 
\begin{cases}
\alpha g^2 \omega^{-5} &\text{ for } \omega>\omega_p,\\
0 &\text{ for } \omega\leq \omega_p,
\end{cases}
\end{equation}
where we take $\alpha=0.0083$ as in \cite{breivik2014approximate}. We relate the peak frequency $\omega_p$ to a friction velocity $u_*$ (see \cite{Holthuijsen2007}) in order to establish $k_M$ as above, and show the results in Figure \ref{fig:Phillips-SD-Comparison}.
Panel (a) shows the Stokes drift near the surface using either the first term in \eqref{eq:Spectral-SD-I-and-II} only (SD I) or both terms in the same equation (SD I+II). The difference between the two is sizable, but strongly dependent on the choice of high-frequency cut-off $k_M.$ This tracks with the very pertinent discussion in \cite{Lenain2020} (therein they report an underestimation of the Stokes drift of up to 50\% from failure to account for high frequencies, but they do not consider the effect of difference harmonics). This difference persists as we descend into the fluid, and a noticeable difference on the order of 1 \% is visible at depths below one peak wavelength in panel (b). Defining the Stokes transport $V_S$ as 
\[ V_S = \int_{-\infty}^0 u_S(z) dz \]
we find nearly a 9\% change in total Stokes transport. If we calculate the Stokes transport only below of the $e$-folding depth of the peak frequency, we still observe a change of approximately 2\% owing to the inclusion of difference harmonic terms (II).

\begin{figure}[h]
\centering
\includegraphics[width=\linewidth]{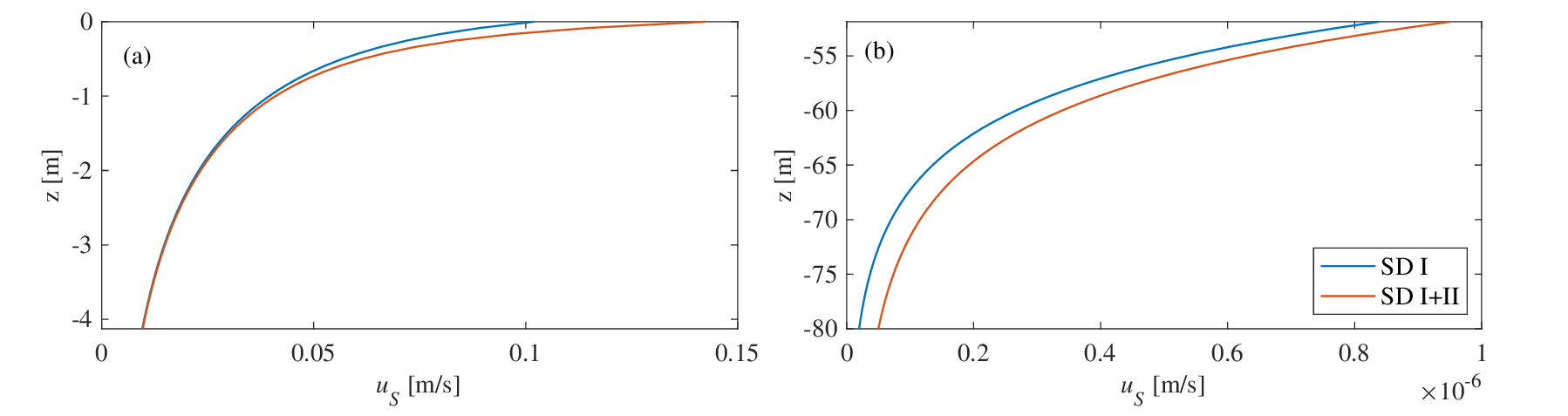}
\caption{Comparison of Stokes drift formulations for a PM spectrum with $U_{10}=10$ m/s. Panel (a) compares the Stokes drift near the surface up to the $e$-folding depth of the peak wavenumber ($k_p=0.1$ 1/m). Panel (b) compares the Stokes drift formulations at depths below one peak wavelength.}
\label{fig:PM-SD-Comparison}
\end{figure}

The Phillips spectrum describes only the high-frequency tail, so it is of interest to consider a more realistic parametric spectrum: we next consider a Pierson-Moskowitz (PM) spectrum, corresponding to a fully developed sea under a constant wind over unlimited fetch. The case shown in Figure \ref{fig:PM-SD-Comparison} corresponds to a wind speeds of $U_{10}=$10 m/s (with corresponding $U_{19.5}\approx 1.026 U_{10}$), and peak frequency $f_p$ of 0.17 Hz. Blue curves show the formula of Kenyon \citeyearpar{Kenyon1969} (corresponding to the first term in \eqref{eq:Spectral-SD-I-and-II}, labelled SD I) and red curves show both terms of \eqref{eq:Spectral-SD-I-and-II} (labelled SD I+II). For reference, we note that the high-frequency cut-off $f_M$ obtained from $k_M$ is 6.3$f_p.$ We note also that it is possible to cut off the low frequencies, and evaluate the outer integrals in \eqref{eq:Spectral-SD-I-and-II} starting from $k_p$ instead of 0 without materially affecting the results, as suggested by \cite{breivik2016stokes,Lenain2020}.

The differences in the formulations are clearly visible, both near the surface and at depth. At the surface, we find increases of surface drift in excess of 40\% when difference harmonic terms are accounted for, reducing to just less than 2\% at half the $e$-folding depth of the peak wavenumber $k_p.$ Interestingly, starker differences reappear at greater depths, as shown in panel (b) which highlights the different asymptotics of the two formulations at depths of more than one peak wavelength. We find that keeping both terms in \eqref{eq:Spectral-SD-I-and-II} leads to an increase in Stokes transport of nearly 3.3 \% for $U_{10}=10$ m/s. The majority of this difference comes from the near-surface transport, with transport below the peak $e$-folding depth accounting for a difference of only 0.1\%. The same analysis can be carried out for JONSWAP spectral shapes, with qualitatively very similar results to those found in Figure \ref{fig:PM-SD-Comparison}.

\begin{figure}
\centering
\includegraphics[width=\linewidth]{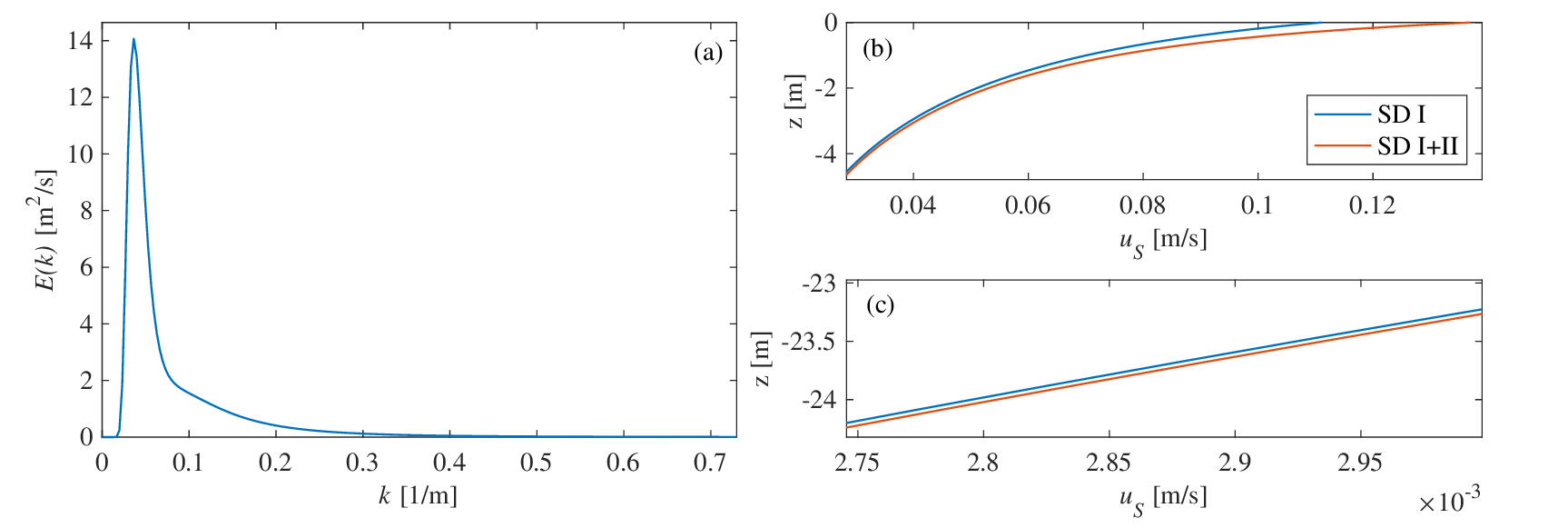}
\caption{Comparison of Stokes drift formulations for an Ochi-Hubble spectrum with $H_s=3$ m. Panel (a) shows the spectral shape as a function of wavenumber $k$, while panels (b) and (c) compare the use of the first term of \eqref{eq:Spectral-SD-I-and-II} (SD I) and both terms of \eqref{eq:Spectral-SD-I-and-II} (SD I+II).}
\label{fig:OH-SD-Comparison}
\end{figure}

We can also compare the Stokes drift formulations of equation \eqref{eq:Spectral-SD-I-and-II} for other spectral shapes, such as swell-wave spectra described by \cite{ochi1976six}. In Figure \ref{fig:OH-SD-Comparison} we show the most probable Ochi-Hubble shape corresponding to a significant wave-height $H_s=3$ m, and the two Stokes drift calculations. The surface drift shows a change in excess of 20 \% when subharmonic terms (II) are included, and the Stokes transport $V_S$ is slightly more than 1\% larger over the entire water column.

\section{Discussion and conclusions}
\label{sec:Conclusions}

The wave induced drift of particles has attracted increasing attention in recent years, particularly in connection with the transport of maritime pollution. Nevertheless, explorations of the underlying hydrodynamics are not widespread. Notable exceptions are the recent articles by  \cite{vandenBremer2016} and the follow-up study by  \cite{VandenBremer2019}, which focused on the transport associated with localised wave groups to second order. Our aim has been to explore particle drift in periodic waves systematically using the first, second, and third order theory, working our way from monochromatic waves to random seas. 

Monochromatic waves -- or progressive Stokes waves -- are extremely well studied, and therefore serve as a useful test case for any development of nonlinear wave theory. They allow us to compare with exact solutions at the surface \citep{Longuet-Higgins1987} as well as high-order approximate solutions in Eulerian \citep{zhao2022stokes} and Lagrangian \citep{Clamond2007} variables. Moreover, these waves can be easily and uniquely defined by specifying a wave height and period, as well as an Eulerian current \citep{Fenton1990}, meaning that waves at various orders of approximation can be directly compared to one another. 

In the absence of an (Eulerian) current, the Eulerian mean flow for such waves vanishes on and below $z=0$, so that the Lagrangian mean velocity is equal to the Stokes drift. Approximating the Stokes drift using the classical formula from linear wave theory shows that this is remarkably accurate, at least when compared to other inviscid, irrotational theoretical results. Some differences do emerge for waves of large steepness, where the Stokes drift formula derived from linear wave theory tends to underestimate drift at the surface and overestimate it at depth. In such cases, we find that the exact results for Lagrangian surface drift by Longuet-Higgins \citeyearpar{Longuet-Higgins1987} as well as the fourth-order, depth-dependent results derived in Lagrangian variables by \cite{blaser2025increased} are approached by the Lagrangian drift obtained from solving the particle trajectory mapping using higher-order solutions. 

With a view towards understanding more complex patterns of waves (and wave groups) on the open sea, we have subsequently studied the unsteady flow associated with bichromatic waves, generated at lowest order by two Fourier harmonics. Unfortunately, very few experiments have been carried out on bichromatic waves, and available results such as those by \cite{westhuis2001experiments} focus on the free surface elevation and the development of instabilities rather than induced flows. Other comparisons of horizontal velocities in bichromatic waves \citep{stansberg2008kinematics} rely on numerical simulations only. The problem is compounded by the fact that bichromatic waves are not themselves well-defined, since the inclusion of higher-order effects leads to a phase shift between the modes. This makes comparisons between different theories fraught with difficulty, and despite the fact that the bichromatic wave train is an exact solution to the 3rd order water wave problem \cite{Mei2005} no exact numerical solutions and attendant kinematics appear to be available in the literature.

That said, our investigations are able to shed some light on the flows induced by bichromatic waves. It is well known that the water column acts like a filter, first suppressing the highest frequencies and filtering out ever lower frequencies with increasing depth. Therefore the particle motion of monochromatic waves is typically considered negligible half a wavelength below the surface. It is this filtering effect on linear, free-wave constituents that leads to the dramatic decrease in Lagrangian drift in the linear theory for bichromatic waves. However, nonlinearity introduces much longer bound-waves which become dominant at depth, as was recognised in the landmark work of \cite{Longuet-Higgins1964}. Beneath the centre of a wave group we find that these can induce sizeable spatially-localised flows opposite the direction of wave propagation. However, when averaged over the period of the bichromatic group the overall effect of higher order terms is to increase the drift in the direction of wave propagation versus linear theory. Third-order contributions lead to additional bound waves, but also to dispersion corrections, which tend to retard the drift slightly.

In the bichromatic case the potential at second order includes new -- difference -- harmonics, in contrast to the monochromatic potential, which is identical at first and second order. This provides an impetus to update the Stokes drift formula, simply by using the second order potential in the Taylor series expansion in place of the first order. The new term -- which captures the Stokes drift of second-order difference harmonics, can make a substantial contribution to the depth dependent drift depending on the wavenumbers involved. This contribution is also captured when simply integrating the particle trajectory ODEs in second- or third-order theory, as well as when using a fourth order Lagrangian drift due to \cite{blaser2025increased}. This corroborates the idea that modifying the Stokes drift formula by the inclusion of difference harmonics can be useful, with the important caveat that all modes (including subharmonics) are assumed in deep water (see below). It is likewise important to emphasise that the new difference harmonic Stokes' drift term itself cannot be directly observed: it is an approximation to the Stokes drift, which, together with any Eulerian mean velocity, makes up the Lagrangian mean velocity. It is the latter which is actually responsible for the motion of particles, and direct measurement of Stokes drift (for example in a laboratory) would require simultaneous knowledge of Eulerian and Lagrangian velocities.

Waves with more than two fundamental harmonics at lowest order become increasingly complex - from an algebraic perspective when deriving an analytical theory, and from a practical perspective when attempting to meaningfully describe the various parameter regimes. As with bichromatic waves, for focused wave groups we see strong localised displacements at depth in the opposite of the wave propagation direction, but find that the averaged drift (over the finite period of the group) remains positive, with slightly larger values for nonlinear theories that include the effects of subharmonics. An important difference between our work and that reported by van den Bremer and coworkers \citep{vandenBremer2016,VandenBremer2019,higgins2020lagrangian} is that the groups considered in our context are periodic, rather than localised. This leads to the qualitatively different conclusions surrounding averaged Lagrangian drift. In particular, for these periodic cases we find (when suitably averaged over a period, or -- in the case of random wave fields -- a sufficiently long time) a net Lagrangian drift in the direction of wave propagation throughout the water column, in contrast to the net displacement opposite the waves found below the transition depth when using a localised group which tends to zero as $x-c_g t \rightarrow \pm \infty$ (for $c_g$ the group velocity).

We further explore the consequences of extending the ubiquitous formulation of \cite{Kenyon1969} by including difference harmonic terms for energy spectra, and obtain significant changes to surface Stokes drift and Stokes transport for several parametric spectral shapes. These changes are strongly dependent on resolving high frequencies (see \cite{Lenain2020}), and it would be interesting to see if such effects can be seen in field experiments or suitable ensemble-averaged flume experiments, or even recovered numerically from suitable Monte-Carlo simulations. 

Throughout, our objective has been to provide a systematic and transparent exposition of drift associated with waves in nonlinear potential flow. To this end numerous simplifications have been made. Noteworthy among these is the neglect of viscous effects at the free surface, which has been known since work of \cite{Longuet-Higgins1953} to play an important role. We have likewise neglected directional effects. Although our methodology can be adapted without change to include wave directional spreading, the additional freedom that this provides makes for a significant additional burden. Nevertheless, directional spreading is key when comparing with real sea states, and it would be of interest to follow in the footsteps of the recent work by \cite{higgins2020lagrangian} and explore these effects. Finally, and  significantly, we have relied on the simplification of infinite water depth. This has allowed us to present numerous simple and compact formulations which are difficult to find in the existing literature, and contributed -- it is hoped -- to a clear presentation. Nevertheless, it is important to note that long bound harmonics may ``feel the bottom" even when the free waves do not, and so the inclusion of finite depth is of central importance. Indeed, finite depth effects can also notably shape the particle drift for monochromatic waves, a subject which has been the subject of exploration from the work of Longuet-Higgins \citeyearpar{Longuet-Higgins1953} to the present day \citep{Grue2017}. We intend to treat finite depth waves in subsequent work.

There are also numerous other cases of interest that should be mentioned. It might be argued that, after Stokes waves the next most-complex -- and already unsteady -- solution is that of standing waves. While some higher-order solutions are available in both Eulerian \citep{Stiassnie1994,Schwartz_Whitney_1981} and Lagrangian \citep{chen2009third} variables, the theory is not nearly as complete as that for periodic, progressive waves. A systematic exploration of the kinematics of standing waves at various orders, in line with that suggested here for progressive waves, would be an interesting area for future work.

Because of our assumptions of infinite water depth for all modes and a vanishing Eulerian mean flow it is quite difficult to compare with experimental results obtained in closed wave flumes, wherein Eulerian mean flows are typically depth-dependent, and therefore not irrotational. Such shear currents and the role of Stokes drift have also been recently investigated in connection with the modified nonlinear Schr\"odinger equation \cite{li2024currents}. Our methodology could be readily adapted to include uniform currents, and so enable a comparison of higher-order Eulerian and Lagrangian kinematics (the latter derived by \cite{chen2012particle}) in this setting. Whether Eulerian mean flows could arise, particularly in connection with amplitude evolution or at higher order, and so alter some of the results discussed in this manuscript, is another interesting question for further study. The possibility of including shear currents also presents a challenge for the future.

The fact that weakly nonlinear theories in Eulerian variables generally make use of Taylor expansion to transfer the boundary conditions to a half-space (and therefore confine the fluid domain to $z\leq 0$) has unfortunate consequences. It has given rise to a significant literature aimed at the problem of evaluating kinematics in the surface zone between the crest and trough level, and particularly at wave crests, beginning with the stretching methods of Wheeler \citeyearpar{wheeler1970method} and continuing to the present day. Recent comparisons of these methods can be found, for example, in Stansberg et al \citeyearpar{stansberg2008kinematics} or Johannessen \citeyearpar{johannessen2010calculations}. We have not resolved the problem of identifying kinematics between crest and trough in Eulerian variables, but we have attempted a theoretically consistent, order-by-order exploration of a variety of waves.

With all of the caveats above, it is important to mention that the drift incorporated into operational models such as WAVEWATCH III is the simplest infinite-depth Stokes drift formulation \citep{Li2016}. This points to a continued need to better understand this formulation -- with all the simplifications it involves -- and how it may be further refined. We hope that the present work presents a modest step in this direction.

\appendix

\section{On the non-uniqueness of Stokes expansion}
\label{app:Non-uniqueness of Stokes exp}

The perturbation expansion associated with Stokes waves is often presented in textbooks as following a set sequence of steps. In fact, numerous choices are made, including whether to add linear terms in $x$ or $t$ to the potential, and how to treat constants appearing in the expansion. A useful discussion of some of these choices is found in \cite[Ch.\ 8]{Svendsen2005}. We remark here on one aspect of this non-uniqueness that has received rather less attention (though see the comment in \cite[Appendix A]{Janssen2009}).

For a chosen linear wave amplitude there is a non-uniqueness in the Stokes' expansion, which makes it necessary to specify the potential explicitly. For example, the following pairs of third-order solutions have the same linear amplitude $a:$
\begin{subequations}
\begin{align}  \label{eq:App-Sol-1a}
\phi(\xi,z) &= \frac{ag}{\omega} e^{kz} \sin(\xi) \\ \label{eq:App-Sol-1b}
\zeta(\xi) &= a \left( 1+ \frac{a^2 k^2}{8} \right) \cos (\xi) + \frac{a^2 k}{2} \cos(2\xi) + \frac{3}{8} a^3 k^2 \cos(3\xi) 
\end{align}
\end{subequations}
and
\begin{subequations}
\begin{align} \label{eq:App-Sol-2a}
\phi(\xi,z) &= \frac{ag}{\omega} \left(1-\frac{a^2 k^2}{4}\right) e^{kz} \sin(\xi) \\ \label{eq:App-Sol-2b}
\zeta(\xi) &= a \left( 1- \frac{a^2 k^2}{8} \right) \cos (\xi) + \frac{a^2 k}{2} \cos(2\xi) + \frac{3}{8} a^3 k^2 \cos(3\xi) 
\end{align}
\end{subequations}
The former \eqref{eq:App-Sol-1a}--\eqref{eq:App-Sol-1b} is given by \cite[(27.25)]{Wehausen1960}, and appears naturally when the third order potential is set to zero. However, at each order solutions to the Laplace equation permit the addition of first-harmonic terms of the form $\sin(\xi)$. The latter \eqref{eq:App-Sol-2a}--\eqref{eq:App-Sol-2b} is the natural form obtained from the Zakharov formulation (see also \cite[(A.16)]{Janssen2009}, \cite[(6.7c)]{Gao2021}). 
Note that this flexibility in the perturbation expansion explains the seeming discrepancy identified by Zhang \& Chen \citeyearpar{Zhang1999} when matching these solutions.

When dealing with monochromatic waves, it is important to note that the solutions \eqref{eq:App-Sol-1a}--\eqref{eq:App-Sol-1b} and \eqref{eq:App-Sol-2a}--\eqref{eq:App-Sol-2b} need to be adjusted when a wave of fixed height is considered. According to these formulations, the height of the wave, defined as $H=\zeta(0)-\zeta(\pi/2)$ is related to the linear amplitude $a$ by either 
\[ H_1 = 2a + a^3 k^2, \quad H_2 = 2a + \frac{1}{2}a^3 k^2,\]
where subscript 1 refers to   \eqref{eq:App-Sol-1a}--\eqref{eq:App-Sol-1b} and subscript 2 to \eqref{eq:App-Sol-2a}--\eqref{eq:App-Sol-2b}. For a prescribed height these expressions can be iteratively inverted to yield
\begin{equation} \label{eq:App-a-H-rel-1-2}
a = \frac{1}{2}H_1 - \frac{1}{16}H_1^3 k^2 + \frac{3}{128}k^4 H_1^5 - \ldots = \frac{1}{2}H_2 - \frac{1}{32}H_2^3 k^2 + \frac{3}{512}k^4 H_2^5 - \ldots.  
\end{equation}
Inserting this into the third order solution recovers the coefficients in terms of $H$ found, for example, in Fenton \cite{Fenton1990}.

\section{Coefficients of the higher-order potential}
\label{app:Coefficients of potential}

We provide here the coefficients of the second-order and third-order contributions to the potential from Section \ref{ssec:Recovery of the third-order solution}. The kernels $\A{i}, \, B^{(i)}$ appearing therein follow the usage of \cite{Krasitskii1994}.

Second-order coefficients are:
\begin{align}
\mathcal{C}^{(2)}_{i+j} &= \frac{1}{\pi} \sqrt{\frac{g}{2 \omega_{i+j}}} \left( \A{1}_{i+j,i,j} - \A{3}_{-i-j,i,j} \right) - \frac{1}{4 \pi^2} \sqrt{\frac{\omega_i}{\omega_j}}|k_j|,\\
\mathcal{C}^{(2)}_{i-j} &= -\frac{1}{\pi} \sqrt{\frac{g}{2 \omega_{i-j}}} \A{2}_{j-i,i,j} + \frac{1}{4 \pi^2} \sqrt{\frac{\omega_i}{\omega_j}}|k_j|.
\end{align}

The third-order coefficients are:
\begin{align} \nonumber
\mathcal{C}^{(3)}_{i+j+k} &= \frac{1}{\pi} \sqrt{\frac{g}{2 \omega_{i+j+k}}} \left[ \B{1}_{i+j+k,i,j,k} - \B{4}_{-i-j-k,i,j,k} \right] - \frac{1}{4 \pi^2} \left[ \sqrt{\frac{\omega_k}{\omega_{i+j}}} |k_i+k_j| \left( \A{1}_{i+j,i,j} - \A{3}_{-i-j,i,j} \right) \right.\\
& \left. + \sqrt{\frac{\omega_{i+j}}{\omega_k}} |k_k| \left( \A{1}_{i+j,i,j} + \A{3}_{-i-j,i,j} \right) \right] - \frac{1}{32 \pi^3} \sqrt{\frac{2 \omega_j \omega_k}{g \omega_i}} \left[ |k_i| \left( |k_i| - |k_i+k_j| - |k_i+k_k| \right) \right], \\  \nonumber
\mathcal{C}^{(3)}_{i-j-k} &= -\frac{1}{\pi} \sqrt{\frac{g}{2 \omega_{i-j-k}}} \B{2}_{j+k-i,i,j,k} + \frac{1}{4 \pi^2} \A{2}_{j-i,i,j} \left[ \sqrt{\frac{\omega_{j-i}}{\omega_k}} |k_k| + \sqrt{\frac{\omega_k}{\omega_{j-i}}}|k_i - k_j| \right] \\
& - \frac{1}{32 \pi^3} \sqrt{\frac{2 \omega_j \omega_k}{g \omega_i}} \left[ |k_i| \left( |k_i| - |k_i-k_j| - |k_i-k_k| \right) \right], \\ \nonumber
\mathcal{C}^{(3)}_{i+j-k} &= - \frac{1}{\pi} \sqrt{\frac{g}{2 \omega_{i+j-k}}} \B{3}_{k-i-j,i,j,k} - \frac{1}{4 \pi^2} \left[ \sqrt{\frac{\omega_k}{\omega_{i+j}}}|k_i+k_j| \left( \A{1}_{i+j,i,j} - \A{3}_{-i-j,i,j} \right) \right. \\
& \left. - \sqrt{\frac{\omega_{i+j}}{\omega_k}} |k_k| \left( \A{1}_{i+j,i,j} - \A{3}_{-i-j,i,j} \right) \right] + \frac{1}{32 \pi^3} \sqrt{\frac{2 \omega_j \omega_k}{g \omega_i}} \left[ |k_i| \left( |k_i| - |k_i+k_j| - |k_i-k_k| \right) \right], \\
\mathcal{C}^{(3)}_{i-j+k} &= - \frac{1}{4 \pi^2} \A{2}_{j-i,i,j} \left[ \sqrt{\frac{\omega_{j-i}}{\omega_k}} |k_k| - \sqrt{\frac{\omega_k}{\omega_{j-i}}} |k_i - k_j| \right] - \frac{1}{32 \pi^3} \sqrt{\frac{2 \omega_j \omega_k}{g \omega_i}} \left[ |k_i| \left( |k_i| - |k_i-k_j| - |k_i+k_k| \right) \right]. 
\end{align}
As previously the shorthand notation $\omega_{i\pm j}$ is used to denote $\omega(k_i \pm k_j) = \sqrt{g |k_i \pm k_j|}$, and similarly for $\omega_{i\pm j \pm k}.$ 

\section{Effects of amplitude evolution on particle kinematics}
\label{app:Amplitude evolution}
In Section \ref{ssec:Constant magnitude approximation} we contend that it is appropriate to neglect the evolution of Fourier amplitudes while calculating particle trajectories and associated drift. This is based on the scale separation between the slow amplitude evolution (taking place on a time-scale of $O(\epsilon^2)$) and the much faster particle motions, whose time-scale is in principle $O(1)$, i.e.\ proportional to the wave period. As we have seen, the motion of an individual particle depends on the wave phase, and so a single estimate of Lagrangian drift is -- in some sense -- an average value only. This is borne out by our methodology for calculating the horizontal drift of particles using 1st, 2nd, and 3rd order theory for monochromatic waves (Figure \ref{fig:Depth-Stokes-drift-monochromatic}), bichromatic waves (Figure \ref{fig:Bichromatic-Stokes-drift-at-depth}) and multichromatic waves (Figure \ref{fig:Random-waves-depth-dependent-drift}).

\begin{figure}[h]
\centering
\includegraphics[width=\linewidth]{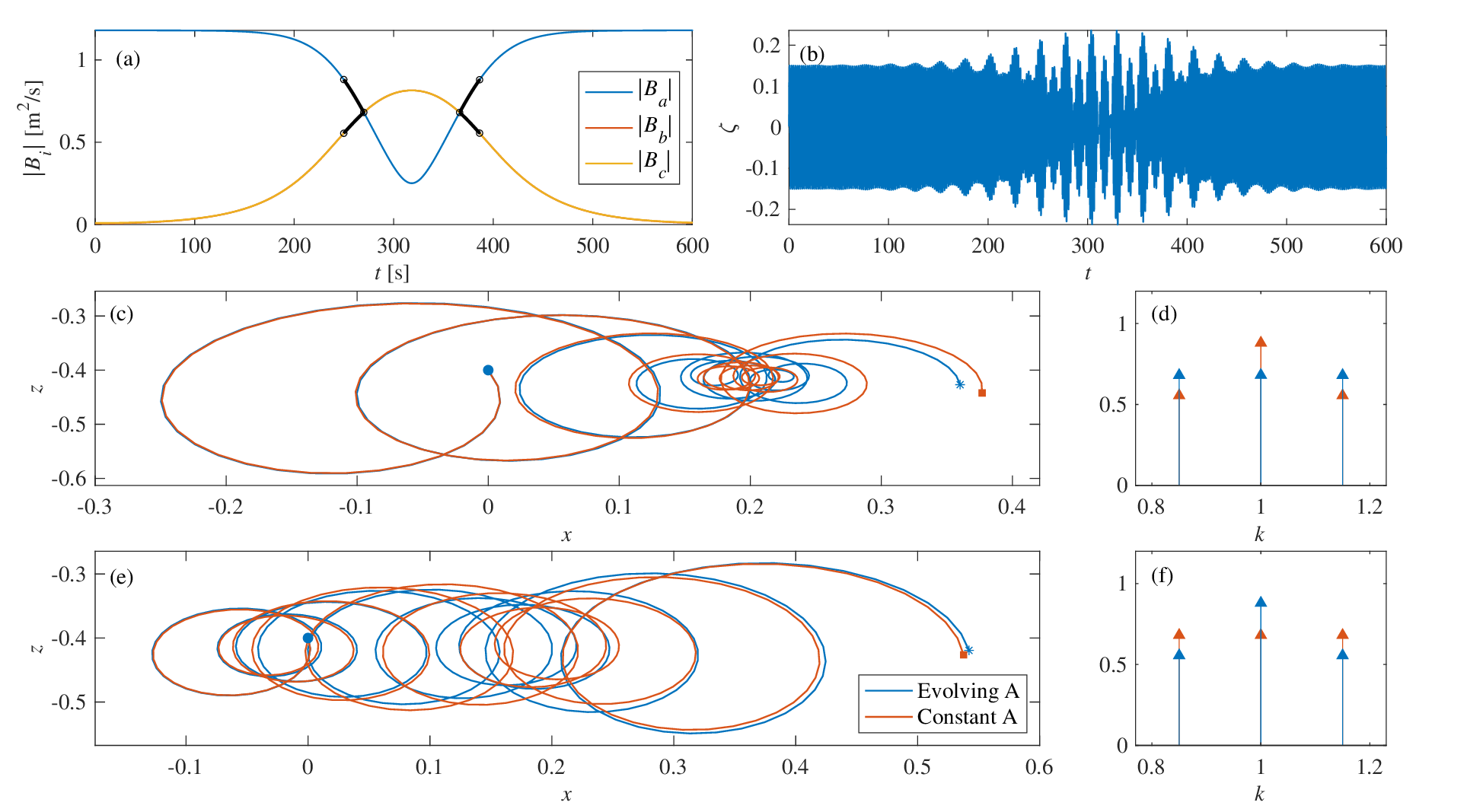}
\caption{Comparison of particle paths with and without time evolution for a modulationally unstable degenerate quartet of waves. Panel (a) shows the evolution of the complex amplitudes over approximately 300 periods $T_a$ of the carrier wave $k_a$ (600 s). Panel (b) shows the corresponding free surface, demonstrating the growth and decay of the modulation. Panel (c) shows the position of a particle initially at $(x_0,z_0)=(0,-0.4)$ over 10 $T_a$ with (blue) and without (red) amplitude evolution. The corresponding evolution of the Fourier amplitudes $|B_i|$ is shown in panel (d), where the initial amplitudes are shown in red, and the amplitudes after $t=10 T_a$ are shown in blue. This period of $10T_a$ is also shown in panel (a) (black curves beginning at $\sim t=250$ s, with start and end points marked with $\circ$.). Panels (e) and (f) are analogous to panels (c) and (d), but capture the transfer of energy from the side-bands to the carrier.}
\label{fig:BFI-PP}
\end{figure}

We explore the role of amplitude evolution for an extreme example in Figure \ref{fig:BFI-PP}, which considers the evolution of a modulationally unstable wave with $k_a=1$ 1/m and (linear) steepness $\epsilon_a=0.15$ together with two initially small disturbances $k_b=k_a+p$ and $k_c=k_a-p$ for $p=0.15.$ For clarity, our comparison excludes 2nd and 3rd order bound modes (see Sec.\ \ref{ssec:Recovery of the third-order solution}), and focuses only on the effects of amplitude evolution.

At the initial time $t=0$ the free surface is essentially monochromatic (Fig.\ \ref{fig:BFI-PP}(b)) and nearly all of the energy is the carrier wave $|B_a|$. Subsequently, instability gives rise to energy exchange between the carrier and the side-bands $|B_b|$ and $|B_c|,$ which can be captured by solving the Zakharov equation \eqref{eq:ZE}, and is shown in Figure \ref{fig:BFI-PP}(a). Note that the characteristic time for energy exchange, $T_a/\epsilon_a^2 \sim 90$ s, where $T_a$ is the carrier period.

Panel (c) of Figure \ref{fig:BFI-PP} shows a particle trajectory at an initial depth of $z_0=-0.4$ m. The particle is tracked for 10$T_a$ starting at time $t\approx 250$ s -- indicated by black curves on the left of panel (a). The amplitude evolution here is such that the energy is being transferred from the carrier to the side-bands, until equipartition is reached in the Fourier amplitudes. The energy transfer is captured in Panel (d), which shows the initial Fourier amplitudes in red (carrier larger than side bands) and the final Fourier amplitudes after 10$T_a$ in blue. In Panel (c) we compute the particle trajectory using both the initial amplitudes only (red curves) and with time-dependent amplitude evolution of Panel (a) (blue curves). Panel (e) shows the corresponding evolution away from equipartition in amplitudes, i.e.\ starting at $|B_a|=|B_b|=|B_c|$ as shown in Panel (f), and following the amplitude evolution for 10$T_a$ along the black curves in Panel (a).

As alluded to earlier, this is an extreme case characterised by a very narrow spectrum and a dramatic spectral evolution from monochromatic waves to a strongly modulated wave train (see Panel (b)). Nevertheless, the amplitude evolution is essentially negligible over a single carrier period $T_a$, and not dramatic even over time-scales of $10 T_a.$ Moreover, any change in particle paths due to amplitude evolution depends both on the phase-position of the particle as well as where we are in the cycle of energy exchange.

\begin{figure}[h]
\centering
\includegraphics[width=\linewidth]{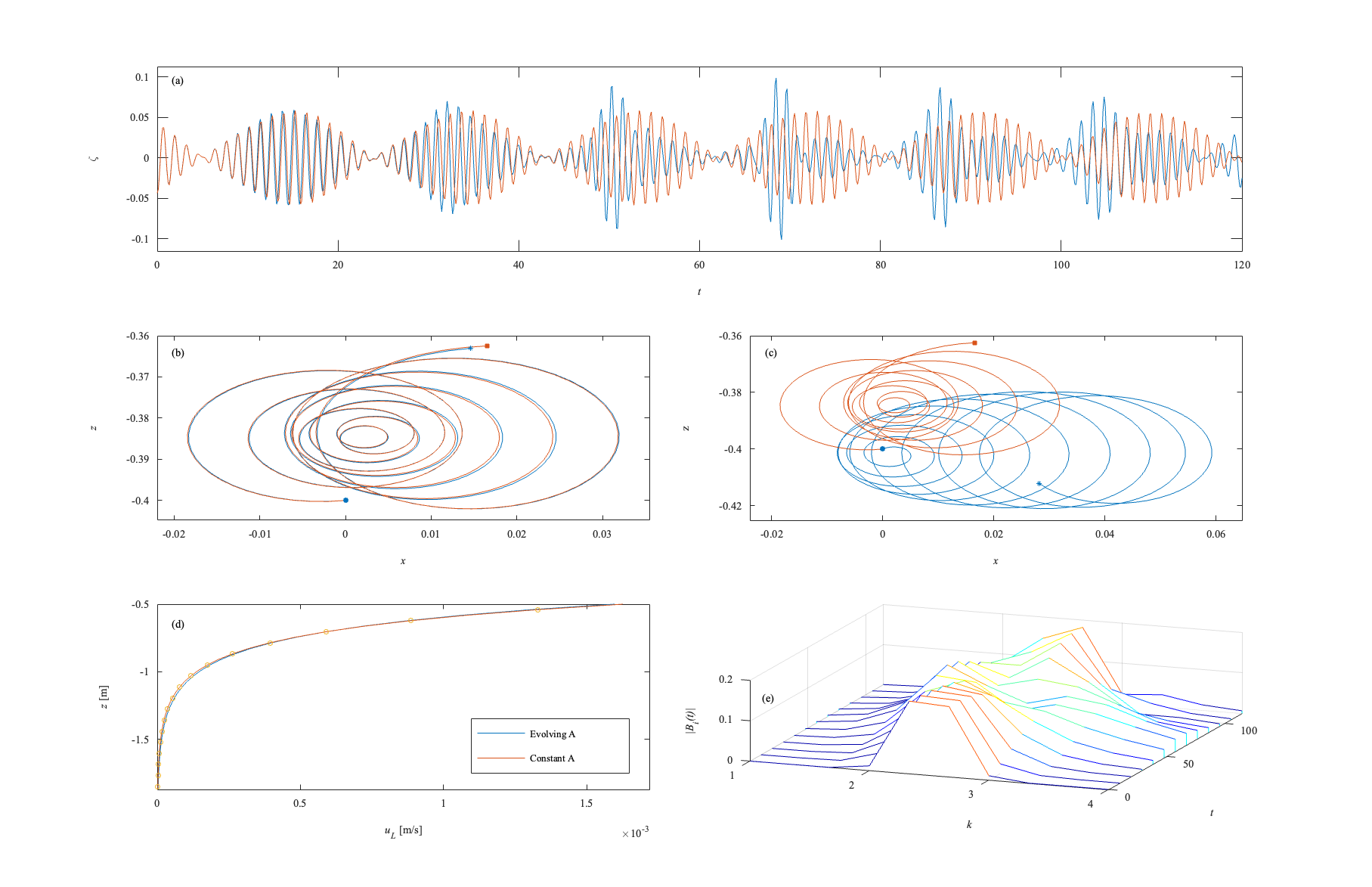}
\caption{Particle trajectories and drift for a constant (red) and evolving (blue) Gaussian amplitude spectrum with $k_p=2.5$ 1/m. The initial spectrum (panel (e), $t=0$) with maximum slope $S=0.15$ is allowed to evolve according to the Zakharov equation, leading to spectral broadening and energy exchange (e). Consequently, the free surface structure is observed to change (a). Particle trajectories below the trough level ($z_0=-0.4$ m) are shown for 10 peak periods in panels (b) and (c), starting at $t=0$ s and $t=70$ s, respectively. The depth-dependent Lagrangian drift, averaged over 20 peak wavelengths, is shown in panel (d), with the lowest order Stokes drift formula (SD I) shown in yellow circles.}
\label{fig:Spectral-PP}
\end{figure}

In Figure \ref{fig:Spectral-PP} we consider a rather milder type of nonlinear interaction (though still significant enough to be interesting), where a Gaussian spectrum (with $k_p=2.5$ 1/m, $\sigma=0.2$ 1/m, and $A=0.04$ m as per \eqref{eq:Gaussian-amplitude-spectrum}) consisting of 10 Fourier modes with maximum initial slope $S=0.15$ is allowed to evolve according to the cubic Zakharov equation \eqref{eq:ZE}. The energy exchange between modes leads to changes in the packet-structure of the free surface (shown in Panel (a)), owing to the spectral broadening which is illustrated in Panel (e), which is solved for approximately $100 T_p$.

Panels (b) and (c) of Figure \ref{fig:Spectral-PP} show particle paths evolving for 10 peak periods starting at $t=0$ and $t=70$ s, respectively. As in Figure \ref{fig:BFI-PP}, blue curves include the amplitude evolution captured by the Zakharov equation while red curves employ constant, initial amplitudes. Unsurprisingly, given the slow rate of spectral change, close to $t=0$ s there is scarce difference in the particle paths over 10 $T_p$ in panel (b). On the other hand, further into the time evolution the wave fields are clearly different, and particle paths cannot be expected to agree.
Despite this, when averaging the drift of particle trajectories over 20 peak wavelengths we find the Lagrangian drift for evolving amplitudes agrees very well with the result for constant amplitudes, shown in Panel (d). In fact, due to the exclusion of bound modes we find that the Lagrangian drift is very close to the linear Stokes drift formula (I) from \eqref{eq:multi-mode-2nd-order-Stokes-drift} -- shown in yellow circles in panel (d).

Recent work by \cite{blaser2025increased} studied the Stokes drift of wave packets undergoing both dispersive focusing and amplitude evolution using a fully nonlinear potential flow solver. Our focus in this appendix is somewhat different from theirs, as we do not concentrate on an isolated wave packet, nor examine the transport for the entire packet; nevertheless, the extent to which a simple cubically nonlinear theory such as that developed here could explain some of the findings from their numerical simulations remains an interesting question for future work.

\end{document}